%% file: ton_caching.tex
\documentclass[journal]{IEEEtran}

\usepackage{amsthm}

\usepackage{graphicx}

\newtheorem{lemma}{Lemma}
\newtheorem{remark}{Remark}

\newtheorem{algorithm}{Algorithm}
\usepackage{citesort}

\newcommand{\be}{\begin{equation}}
\newcommand{\ee}{\end{equation}}
\newcommand{\bea}{\begin{eqnarray}}
\newcommand{\eea}{\end{eqnarray}}
\newcommand{\bw}{\begin{eqnarray*}}
\newcommand{\ew}{\end{eqnarray*}}

\newcommand{\D}{\displaystyle}

\usepackage{amssymb,amsmath}
\usepackage{amsmath}
\usepackage{latexsym}
\usepackage{graphicx}
\usepackage{graphics}
\usepackage{color}

\input{psfig.sty}
\usepackage{psfig}

\usepackage{algorithm,algorithmic}

\IEEEoverridecommandlockouts

\begin{document}

\title{Sprout:  A functional caching approach to minimize service latency in erasure-coded storage}
\author{Vaneet Aggarwal,  Yih-Farn R. Chen,  Tian Lan, and Yu Xiang\thanks{The author names are written in an alphabetical order. V. Aggarwal is with the School of IE, Purdue University, West Lafayette, IN 47907, USA, email: vaneet@purdue.edu. Y. R. Chen and Y. Xiang are with AT$\&$T Labs-Research, Bedminster, NJ 07921, USA, email: \{chen,yxiang\}@research.att.com. T. Lan is with the Department of ECE, George Washington University, DC 20052, USA, email: tlan@gwu.edu
		
		This work was supported in part by the National. Science Foundation under grant CNS-1618335. This work was presented in part in Proc. IEEE ICDCS 2016 \cite{7536586}.}
}

\maketitle
\begin{abstract}
Modern distributed storage systems often use erasure codes to protect against disk and node failures to increase reliability, while trying to meet the latency requirements of the applications and clients. Storage systems may have caches at the proxy or client ends in order to reduce the latency. In this paper, we consider a novel caching framework with erasure code called {\em functional caching}. Functional Caching involves using erasure-coded chunks in the cache such that the code formed by the chunks in storage nodes and cache combined are maximal-distance-separable (MDS) erasure codes. Based on the arrival rates of different files, placement of file chunks on the servers, and service time distribution of storage servers, an optimal functional caching placement and the access probabilities of the file request from different disks are considered. The proposed algorithm gives significant latency improvement in both simulations and a prototyped solution in an open-source, cloud storage deployment.
\end{abstract}
\input{Intro}

\input{related}

\input{System}

\input{Online}

\input{Opti}
\input{Algo}
\input{Imple2}



\section{Conclusions}
In this paper, we propose functional caching, a novel approach to create erasure-coded cache chunks that maintain MDS code property along with existing data chunks. It outperforms exact caching schemes and provides a higher degree of freedom in file access and request scheduling. We quantify an upper bound on the mean service latency in closed-form for erasure-coded storage systems with functional caching, for arbitrary chunk placement and service time distributions. A cache optimization problem is formulated and solved using an efficient heuristic algorithm. Numerical results and prototype in an open-source cloud storage validate significant service latency reduction using functional caching.

 This paper assumes that a rate monitoring/prediction oracle (e.g., an online predictive model or a simple sliding-window-based method) is available to detect the rate changes. Finding a robust algorithm that can automatically adjust to such changes is an open problem and will be considered as future work. 



\bibliographystyle{IEEEtran}
\bibliography{cache,allstorage,Tian,ref_Tian2,ref_Tian3,Vaneet_cloud,Tian_rest,yu}
\input{bio}

\appendices
\input{testbed}
\input{apdx}

\end{document}

%% file: Intro.tex
\section{Introduction}

Erasure coding has seen itself quickly emerged as a promising technique to reduce the storage cost for a given reliability as compared to fully-replicated systems \cite{2015_1,Dimakis:10}. It has been widely adopted in modern storage systems by companies like Facebook \cite{Sathiamoorthy13}, Microsoft \cite{Asure14} and Google \cite{Fikes10}. In these systems, the rapid growth of data traffic such as those generated by online video streaming, Big Data analytics, social networking and E-commerce activities has put a significant burden on the underlying networks of datacenter storage systems. Many researchers have begun to focus on latency analysis in erasure coded storage systems \cite{ISIT:12,Joshi:13,CS14,MDS-Queue,Xiang:2014:Sigmetrics:2014,Yu_TON,Jingxian,aggarwal2017taming} and to investigate algorithms for joint latency optimization and resource management \cite{Yu_TON,Yu-ICDCS,aggarwal2017taming,Yu-CCGRID,YuTCC17,Yu-TNSM17}.

\begin{figure}[!thbp]
\vspace{-2mm}
\begin{center}
\scalebox{0.2}{\fbox{\includegraphics[draft=false]{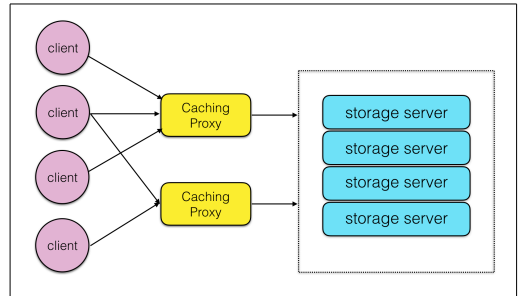}}}
\vspace{-3mm}
\caption{Caching proxies for video storage to reduce the latency of client accesses for popular Video-on-Demand content }
\label{fig:proxy}
\end{center}
\vspace{-.25in}
\end{figure}

Historically, a key solution to relieve this traffic burden is caching \cite{td_cache}. By storing chunks of popular data at different locations closer to end-users, caching can greatly reduce congestion in the network and improve service delay for processing file requests. For example, Figure \ref{fig:proxy} shows a typical video storage architecture with video proxies and multiple video clients.  It is very common for 20\% of the video content to be accessed 80\% of the time, so caching popular content at proxies significantly reduces the overall latency on the client side.

However, caching for datacenters where the files are encoded with erasure codes gives rise to new challenges. The current results fall short of addressing the impact of erasure coding on latency and thus fail to providing insights on the optimal caching policy. First, using an $(n,k)$ maximum-distance-separable (MDS) erasure code, a file is encoded into $n$ chunks and can be recovered from any subset of $k$ distinct chunks. Thus, file access latency in such a system is determined by the delay to access file chunks on hot storage nodes with the slowest performance. Significant latency reduction can be achieved by caching only a few hot chunks of each file (and therefore alleviating system performance bottlenecks), whereas caching additional chunks or even complete files (e.g., \cite{Nadgowda2014,Chang:2008:BDS:1365815.1365816,Zhu:2004:PSP:1006209.1006221,naik2015read}) only has diminishing benefits. It is an open problem to design a caching policy that optimally apportions limited cache capacity among all files in an erasure coded storage to minimize overall access latency.

More importantly, caching the most popular data chunks is often optimal because the cache-miss rate and the resulting network load are proportional to each other. However, this may not be true for an erasure-coded storage, where cached chunks need not be identical to the chunks transferred from storage nodes. More precisely, leveraging the existing erasure coding, 
a function of the data chunks can be computed and cached, so that the constructed new chunks (i.e., $d$), along with the existing chunks, satisfy the property of being a new MDS code. It effectively expands the existing $(n,k)$ code to an $(n+d,k)$, which leads to much lower access latency \cite{Yu_TON}.





In this paper, we propose a new {\em functional caching} approach called \textit{Sprout} that can efficiently capitalize on existing file coding in erasure-coded storage systems. In contrast to exact caching that stores $d$ chunks identical to original copies, our functional caching approach forms $d$ new data chunks, which together with the existing $n$ chunks satisfy the property of being an $(n+d,k)$ MDS code. Thus, the file can now be recovered from any $k$ out of $n+d$ chunks (rather than $k$ out of $n$ under exact caching), effectively extending coding redundancy, as well as system diversity for scheduling file access requests. The proposed functional caching approach saves latency due to more flexibility to obtain $k-d$ chunks from the storage system at a very minimal additional computational cost of creating the coded cached chunks. {To the best of our knowledge, this is the first work studying functional caching for erasure-coded storage and proposing an analytical framework for cache optimization.

Most of the caching strategies in today's data centers cache complete files \cite{software-defined-caching-managing-caches-in-multi-tenant-data-centers,Hong:2013:UMI:2523616.2525970,Cheng:2015:IOC:2741948.2741967}. This flexibility without erasure coding has been recently explored in  \cite{Wang:2015:OCP:2780720.2781435}. Partial number of chunks in  cache give more flexibility as compared to caching the entire file. This is because the connection from some servers may be better than others and caching a smaller number of chunks can aid avoiding the bottleneck. The proposed optimization includes as a special case complete file caching and thus the performance can be no worse than caching the entire file in the cache. This paper gives a novel approach, functional caching, which is an efficient approach to cache partial chunks since any $k-d$ of the $n$ servers can be used thus increasing flexibility as compared to using any $k-d$ of the remaining $n-d$ servers when $d$ exact chunks are copied in the cache. This additional flexibility shows that the latency with functional caching is no higher than the strategy where part of the chunks on the servers are cached as such.}

While quantifying service latency in erasure-coded storage systems is an open problem, we generalize previous results on probabilistic scheduling policy \cite{Yu_TON,Xiang:2014:Sigmetrics:2014} that distributes the file requests to cache and storage nodes with optimized probabilities, and derive a closed-form upper bound on mean service latency for the proposed functional caching approach. The latency bound is obtained using order-statistic \cite{td_os} analysis and it works on erasure-coded storage systems with arbitrary cache size and data chunk placement. 

This analytical latency model for functional caching enables us to formulate a cache-content optimization problem. This problem is  an integer optimization problem, which is very difficult to solve. Towards this end, for given data chunk placement and file request arrival rates, we propose a heuristic algorithm that iteratively identifies files whose service latency benefits most from caching and constructs new functional data chunks until the cache is filled up. The algorithm can be efficiently computed to allow online cache optimization and management with time-varying arrival rates.

The proposed algorithm is an iterative algorithm, which converges within a few iterations in our conducted experiments and it was validated by the numerical results. For 1000 files, we find that the algorithm converges within 20 iterations. The file latency decreases as a convex function as the cache size increases thus showing diminishing returns for the increase in cache size. We also find that it is suboptimal in general to have all $k$ chunks of an $(n,k)$ coded file in the cache. Further, the cache placement depends heavily on the file arrival rates, storage server service time distributions, as well as the content placement on the files. If a high arrival-rate file is placed on servers which have a less overall load, this file may not have any contribution in the cache. Thus, the proposed algorithm accounts for all the system parameters to effectively find the cache placement. {The proposed algorithm is prototyped using {\em Ceph}, an open-source erasure-coded storage system \cite{Weil} and tested on a real-world storage testbed with an emulation of real storage workload.} We find that caching helps improve the latency significantly.

The key contributions of our paper include:
\begin{itemize}
\item We propose a new functional caching scheme that leverages existing file coding in erasure-coded storage systems, and quantify its service latency through probabilistic request scheduling.
\item Based on the service latency analysis, an optimization problem is formulated which minimizes the average latency of the files. This problem has integer constraints due to the integer number of chunks in the cache.
\item An iterative algorithm is developed to optimize cache content. The proposed algorithm takes file placement, service time distributions, and file arrival rates into account to find the cache placement which optimizes the service latency.
\item The simulation results show that the proposed algorithm converges within a few iterations.
\item {The prototype implementation of the functional caching scheme and the cache optimization algorithm using {\em Ceph} are used to validate significant latency reduction on a real-world storage testbed. As compared to the Ceph's LRU (least recently used) caching algorithm, the algorithm proposed in this paper reduces latency by 24.93\% on an average for all tested workloads in the prototype implementation. }
\end{itemize}
The remainder of this paper is organized as follows. Section II provides related work for this paper. In Section III, we describe the system model used in the paper with a description of  functional caching. Section IV formulates the cache optimization problem and develops an iterative algorithmic solution. Prototype and evaluation results are included in Section V. Section VI concludes the paper.

%% file: related.tex
\section{Related Work}

Quantifying exact latency for an erasure-coded storage system is an open problem. Recently, there has been a number of attempts at finding latency bounds for an erasure-coded storage system \cite{ISIT:12,Joshi:13,MDS-Queue,CS14,Xiang:2014:Sigmetrics:2014,Yu_TON,Yu-ICDCS,Yu-CCGRID,7164290,7541297,Joshi:2015:QRL:2825236.2825258}. In this paper, we utilize the probabilistic scheduling policy developed in \cite{Xiang:2014:Sigmetrics:2014,Yu_TON} and analyze the impact of caching on the service latency of erasure-coded storage. {Even though exact characterization of latency is open, probabilistic scheduling has been shown to be optimal for latency tail index, when the file size is Pareto distributed \cite{DBLP:journals/corr/AggarwalL16}.}

Caches are a critical resource in data centers; however, there is { little } work on  caching for erasure coded distributed storage. The problem of online cache management (i.e., decisions for evicting a data chunk currently in the cache to make room for a new chunk) has been studied for networks with distributed caches \cite{td_caches_1,td_caches_2}.  Cache replacement strategy called LRU (Least Recently Used) is widely used
in managing buffer cache  due to its simplicity \cite{1453496,Megiddo:2003:ASL:1090694.1090708,Zhou:2001:MRA:647055.715773,Pelc07,td_lru}. Recently, a steady-state characterization of various cache policies is developed in \cite{td_steady}, and new coded caching schemes to enable multicast opportunities for bandwidth minimization are proposed in \cite{td_cache,td_coded}.   { Recently, caching in erasure-coded storage has been studied \cite{rashmi2016ec} where the cache stores the files in their entirety. In contrast, this paper allows for partial chunks, which are functionally different from those stored on the servers. }

Much of the work on caching in data centers is focused on specialized application caches, such as Facebook's photo-serving
stack \cite{Huang:2013:AFP:2517349.2522722}, Facebook's social graph store \cite{180185}, memcached
\cite{Fitzpatrick:2004:DCM:1012889.1012894}, or explicit cloud caching services \cite{Chockler:2010:DCC:1859184.1859190,6097162}. However, in the presence of coding, new challenges arise. First, the chunks are distributed over multiple servers and a part of the chunks can be in the cache. Thus, it is not necessary for the complete file to be in the cache. Second, the latency calculation for a file depends on the placement of the files and the request of the files from $k$ out of $n$ servers. In this paper, we deal with these challenges to consider a novel caching strategy. To the best of our knowledge, this is the first work on caching with erasure coded files on distributed storage servers accounting for the latency in file retrievals, based on the estimated arrival rates.

Coded caching for a single server with multi-cast link to different users has been considered in \cite{DBLP:journals/corr/NiesenM14,6875212,6763007,6849235}. This does not account for multiple distributed storage servers and latency to get the content. An extension of the approach to distributed storage systems is considered recently in \cite{DBLP:journals/corr/ShariatpanahiMK15,luo2016coded}, where  multiple cache-enabled clients connected to multiple servers through an intermediate network. However, the impact of coding on the servers, and limited service rate of different servers is not taken into account. The key idea in this set of works uses a coded version of different files in the cache which helps in the case when users request different files with certain probabilities. The gain of the approach is due to the model where a message from the server can be received at multiple nodes and thus combined with coded content in the cache, one chunk from the server can help give a chunk for different files at different clients. In this paper, we do not have multicast links to different clients and thus coding across files in the cache is not used. 

The functional repair was introduced in \cite{Dimakis:10} for repairing a failed chunk with a new chunk such that the storage code satisfies the same properties even after repair. Thus, the replaced content can be different. In this paper, we use functional caching to have chunks in the cache such that the file can be recovered from any of $k$ chunks from a combination of disk and cache contents where $(n,k)$ code is used for the files in the storage system. 

%% file: System.tex
\section{System Model}

We consider a distributed storage system consisting of $m$ heterogeneous storage nodes, denoted by $\mathcal{M}=\{1,2,\ldots,m\}$. To distributively store a set of $r$ files, indexed by $i=1,\ldots,r$, we partition each file $i$ into $k_i$ fixed-size chunks\footnote{While we make the assumption of fixed chunk size here to simplify the problem formulation, all results in this paper can be easily extended to variable chunk sizes. Nevertheless, fixed chunk sizes are indeed used by many existing storage systems \cite{DPR04,AJX05,LC02}.} and then encode it using an $(n_i,k_i)$ MDS erasure code to generate $n_i$ distinct chunks of the same size for file $i$. The encoded chunks are stored on the disks of $n_i$ distinct storage nodes. A set $\mathcal{S}_i$ of storage nodes, satisfying $\mathcal{S}_i\subseteq\mathcal{M}$ and $n_i=|\mathcal{S}_i|$ is used to store file $i$. Therefore, each chunk is placed on a different node to provide high reliability in the event of node or network failures. The use of $(n_i,k_i)$ MDS erasure code allows the file to be reconstructed from any subset of $k_i$-out-of-$n_i$ chunks, whereas it also introduces a redundancy factor of $n_i/k_i$.

The files are accessed { (to be read in their entirety)} by compute servers located in the same datacenter. A networked cache of size $C$ is available at each compute server to store a limited number of chunks of the $r$ files in its cache memory. File access requests are modeled by a non-homogenous Poisson process. We make the assumption of time-scale separation, such that system service time is divided into multiple bins, each with different request arrival rates, while the arrival rates within each bin remain stationary. { This model allows us to analyze cache placement and latency in steady-state for each time bin, and by varying arrival rates for different time bins, to also take into account time-varying service rates during busy and off-peak hours. } Let $\lambda_{i,j,t}$ be the arrival rate of file-$i$ requests at compute server $j$ in time bin $t$. { These arrival rates $\lambda_{i,j,t}$ can be estimated using online predictive models \cite{shen2000predictive} or a simple sliding-window-based method, which continuously measures the average request arrival and introduces a new time bin if the arrival rates vary sufficiently.} It is easy to see that estimating arrival rates using a small window is prone to the dynamics of stochastic request arrivals. However, a large window size introduces a low-pass filtering effect, causing higher delay and insensitivity in detecting rate changes. We also note that more advanced methods such as \cite{rate_est1,rate_est2} can be used to estimate arrival rate changes in non-homogeneous processes. Since each cache serves a single compute server, we consider a separate optimization for each cache and suppress server index $j$ in the notations. Let $d_i\le k_i$ (chunks) be the size of cache memory allocated to storing file $i$ chunks. These chunks in  the cache memory can be both prefetched in an offline fashion during a placement phase \cite{td_cache} (during hours of low workload) and updated on the fly when a file $i$ request is processed by the system.

{\noindent \bf Functional Caching.} { Under functional caching, $d_i$ new coded data chunks of file $i$ are constructed and cached, so that along with the existing $n_i$ chunks satisfy the property of being an $(n_i+d_i,k_i)$ MDS code. In this paper, we use Reed-Solomon codes \cite{RS} to generate MDS codes for arbitrary $(n_i+k_i,k_i)$. Thus a subset of the encoded chunks generated using these codes satisfy the desired property for functional caching.} More precisely, for given erasure coding and chunk placement on storage nodes and cache, a request to access file $i$ can be processed using $d_i$ cached chunks in conjunction with $k_i-d_i$ chunks on distinct storage nodes. After each file request arrives at the storage system, we model this by treating the file request as a {\em batch} of $k_i-d_i$ chunk requests that are forwarded to appropriate storage nodes, as well as $d_i$ chunk requests that are processed by the cache. Each storage node buffers requests in a common queue of infinite capacity and process them in a FIFO manner. The file request is served when all $k_i$ chunk requests are processed. Further, we consider chunk service time ${\bf X}_j$ of node $j$ with {\em arbitrary distributions}, whose statistics can be inferred from existing work on network delay \cite{AY11,WK} and file-size distribution \cite{D11,PT12}. 

\input{example}
{

In order to have an $(n,k)$ coded file in the storage server, we can construct chunks by using an $(n+k,k)$ MDS code, where $n$ chunks are stored in the storage server. The remaining $k$ out of the $n+k$ coded chunks are assigned to be in part in the cache based on the contents of the file in the cache. Thus, irrespective of the value of $d\le k$, we ascertain that $(n+d,k)$ code, formed with $n$ coded chunks in the storage server and $d$ coded chunks in the cache will be MDS. }

%% file: example.tex

\begin{figure}[!thbp]
\vspace{-2mm}
\begin{center}
\includegraphics[trim=1in 1.1in 1.2in .5in, clip,width=.47\textwidth,draft=false]{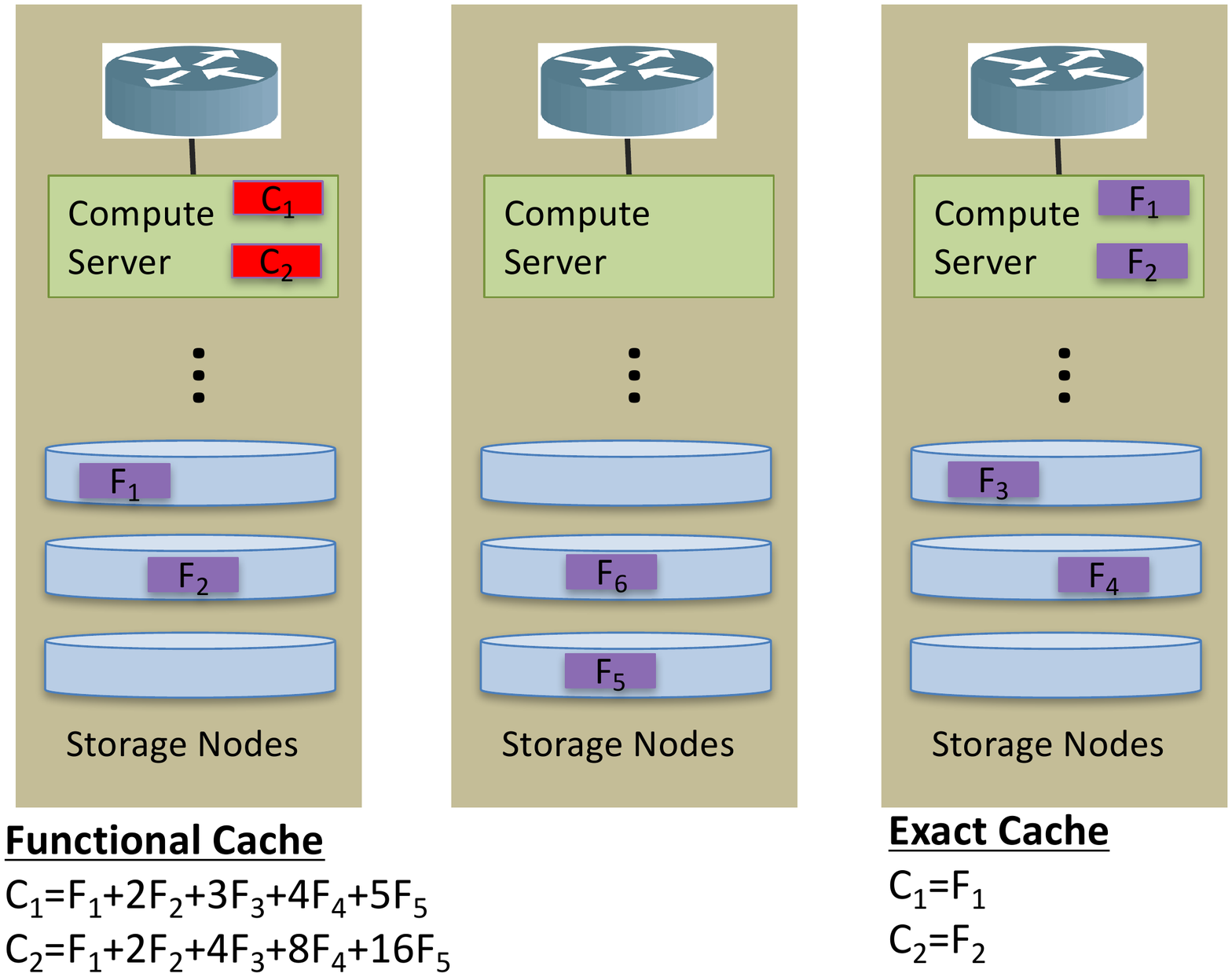}
\vspace{-2mm}
\caption{An illustration of functional caching and exact caching in an erasure-coded storage system with one file using a $(6,5)$ erasure code. The file is split into $k_i=5$ chunks, denoted by $A_1,A_2,A_3,A_4, A_5$, and then linearly encoded to generate $n_i=6$ coded chunks $F_1=A_1$, $F_2=A_2$, $F_3=A_3$, $F_4=A_4$, $F_5=A_5$, and $F_6=A_1+A_2+A_3+A_4+A_5$.
}
\label{fig:example}
\end{center}
\end{figure}

{\noindent \bf An Illustrative Example.}  Consider a datacenter storing a single file using a $(6,5)$ MDS code. The file is split into $k_i=5$ chunks, denoted by $A_1,A_2,A_3,A_4,A_5$, and then linearly encoded to generate $n_i=6$ coded chunks $F_1=A_1$, $F_2=A_2$, $F_3=A_3$, $F_4=A_4$, $F_5=A_5$ and $F_6=A_1+A_2+A_3+A_4+A_5$ in a  finite field of order at-least 5. { In this example, we compare the latency of 2 different cache policies: (i) Exact caching in Rack 3 that stores chunks $F_1$ and $F_2$, and (ii) Functional caching in Rack 1 that stores 2 coded chunks $C_1$ and $C_2$.

Due to the use of $(6,5)$ MDS code, the file can be reconstructed using any $5$ out of $6$ chunks.} Two compute servers in the datacenter access this file and each is equipped with a cache of size $C=2$ chunks as depicted in Figure \ref{fig:example}. The compute server on the right employs an exact caching scheme and stores chunks $F_1,F_2$ in the cache memory. Thus, 3 out of 4 remaining chunks (i.e., $F_3$, $F_4$ , $F_5$ or $F_6$) must be retrieved to access the file. Chunks $F_1,F_2$ and thus their host nodes will not be selected for processing the requests even if the servers in Rack 1 are least congested. Since file access latency is determined by the time required to retrieve all 5 selected chunks, caching $F_1,F_2$ may not necessarily reduce file access latency if other chunks in Rack 2 and 3 are currently the performance bottleneck.

{We show that lower latency can always be achieved in functional caching by exploiting the erasure code structure. More specifically, the compute server on the left generates $d_i=2$ new coded chunks, i.e., $C_1=A_1+2A_2+3A_3+4A_4+5A_5$ and $C_2=A_1+2A_2+4A_3+8A_4+16A_5$, and saves them in its cache memory. It is easy to see that chunks $F_1,\ldots,F_6$ and $C_1,C_2$ now form a $(8,5)$ erasure code. Thus, the file can be retrieved by accessing $C_1,C_2$ in the cache together with any 3 out of 6 chunks from $F_1,\ldots,F_6$. This allows an optimal request scheduling mechanism to select the least busy chunks/nodes among all 5 possible candidates in the system so that the service latency is determined by the best 3 storage nodes with minimum queuing delay.} 
{In order words, the coded chunks $C_1,C_2$ can always be used to replace the slowest chunks required under exact caching policy, resulting in smaller latency. Consider file access latency, which is determined by the time required to access any $5$ chunks on the storage servers. Under an exact caching of file copies $F_1$ and $F_2$, any request to access the file must select 3 other files chunks hosted by storage servers. Thus, access latency in this case is determined by the minimum time to retrieve any 3 chunks out of $F_3$, $F_4$, $F_5$ and $F_6$. In contrast, under our proposed functional caching policy, given the availability of coded chunks $C_1$ and $C_2$ in cache, any 2 chunks from $F_1$, $F_2$, $F_3$, $F_4$, $F_5$ and $F_6$ is sufficient for recovery, which leads to smaller latency if nodes storing $F_1$ and $F_2$ are less congested and faster to access. This approach effectively extends to an $(8,5)$ erasure code and guarantees lower access latency and also enables better load balancing due to a higher degree of flexibility in request scheduling and chunk selection. 

%% file: Online.tex

In the proposed system, the placement in each time-bin is decided based on the predicted arrival rates in the time bin. The time bin can either be a fixed time or dynamic based on a significant change of the predicted arrival rates. At the start of the time-bin, new cache placement is found using the optimized algorithm. For each file that has a lower number of chunks in the new time bin, the decreased contents are removed from the cache. For the files for which the cache contents increase in a time bin, we wait for the file to be accessed. When the file is accessed,  the file contents are gathered and the required new chunks are generated to be placed in the cache. Thus, the change of cache content does not place any additional network overhead and the cache contents of a file are added only when it is first accessed in the new time bin. This process can be further improved improving latency till convergence to the new cache content in the new time bin by not letting all the chunks which have to be removed all simultaneously but removing as needed based on the added chunks.

%% file: Opti.tex
\section{Optimized Distributed Storage with Cache}

In this section, we quantify the mean service latency for file access with functional caching. The result enables us to formulate a cache optimization problem for latency minimization and develop an efficient algorithm solution.

\subsection{Formulation of Optimization}

At time $t$, we consider the cache optimization problem, which decides the optimal number $d_{i,t}$ of file-$i$ chunks to store in the cache memory, satisfying cache capacity constraint $\sum_{i=1}^rd_{i,t}\le C$, in order to minimize mean service latency of all files. Under functional caching, each file-$i$ request is served by accessing $d_{i,t}$ chunks in the cache, along with $k_i-d_{i,t}$ distinct chunks that are selected from $n_i$ storage nodes. Thus, the latency to access file $i$ under functional caching is determined by the maximum processing (queuing) delay of the selected $k_i-d_{i,t}$ storage nodes. Quantifying service latency in such erasure-coded system is an open problem. In this paper, we use probabilistic scheduling proposed in \cite{Yu_TON} to derive an upper bound on the average file latency.

The key idea is to forward each file-$i$ request to a set of $k_i-d_{i,t}$ storage nodes (denoted by $\mathcal{A}_{i,t}\subseteq \mathcal{S}_i$) with some predetermined probabilities $\{\pi_{i,j,t}\in [0,1], \forall i,j,t\}$ for $j\in\mathcal{A}_{i,t}$. Each node then manages a local queue and process chunk requests with service rate $\mu_j$. While the local queues are not independent due to coupled request arrivals, we can leverage order statistic analysis to derive an upper bound of mean service latency in closed-form \cite{Yu_TON}. The result is then optimized over probabilities $\pi_{i,j,t}$ to obtain the tightest bound. Let ${\bf Q}_{j,t}$ be the (random) waiting time a chunk request spends in the queue of node $j$ in time-bin $t$. Using the functional caching approach, requests of file $i$ see mean latency $\bar{T}_{i,t}$ given by
\vspace{-.1in}\begin{eqnarray}
\D \bar{T}_{i,t}= \mathbb{E}\left[\mathbb{E}_{\mathcal{A}_{i,t}} \left(\max_{ j\in \mathcal{A}_{i,t}} \{ {\bf Q}_{j,t} \} \right) \right], \label{eq:T_bar}
\end{eqnarray}
where the first expectation is taken over system queuing dynamics and the second expectation is taken over random dispatch decisions $\mathcal{A}_{i,t}$. We note that queuing delay ${\bf Q}_{j,t}$'s are dependent due to coupled request arrivals. Therefore, an exact queuing analysis of the system is intractable. We use the technique in \cite{Yu_TON} to derive an upper bound of (\ref{eq:T_bar}).


We denote ${\bf X}_j$ as the service time per chunk at node $j$, which has an arbitrary distribution satisfying finite mean $\mathbb{E}[{\bf X}_j]=1/\mu_j$, variance $\mathbb{E}[{\bf X}_j^2]-\mathbb{E}[{\bf X}_j]^2=\sigma^2_j$, second moment $\mathbb{E}[{\bf X}_j^2]=\Gamma_j^2$, and third moment $\mathbb{E}[{\bf X}_j^3]=\hat{\Gamma}^3_j$. These statistics can be readily inferred from existing work on network delay \cite{AY11,WK} and file-size distribution \cite{D11,PT12}. Following \cite{Yu_TON}, an upper bound on the expected latency is given as follows.

\begin{lemma} \label{th:lemma_1}
The expected latency $\bar{T}_{i,t}$ of file $i$    in time-bin $t$ under probabilistic scheduling is upper bounded by $\bar{U}_{i,t}$, given by
\vspace{-.1in}
\begin{eqnarray}
\D & \bar{U}_{i,t} & = \min_{z_{i,t}\in \mathbb{R}} \left\{ z_{i,t}+\sum_{j\in \mathcal{S}_{i,t}} \frac{\pi_{i,j,t}}{2}  \left(\mathbb{E}[{\bf Q}_{j,t}] -z_{i,t} \right) \right.  \label{eq:lemma1} \nonumber\\
& & \left.+ \sum_{j\in \mathcal{S}_{i,t}} \frac{\pi_{i,j,t}}{2} \left[ \sqrt{(\mathbb{E}[{\bf Q}_{j,t}]-z_{i,t})^2+{\rm Var}[{\bf Q}_{j,t}]}\right] \right\},
\end{eqnarray}
where
\vspace{-.2in}\begin{eqnarray}
\mathbb{E}[  {\bf Q}_{j,t}] =  \frac{1}{\mu_j} + \frac{ \Lambda_{j,t} \Gamma_j^2 }{2(1- \rho_{j,t})}, \label{eq:lemma2_1}
\end{eqnarray}
\vspace{-0.2in}
\begin{eqnarray}
{\rm Var}[ {\bf Q}_{j,t}] =\sigma_j^2+\frac{ \Lambda_{j,t} \hat{\Gamma}_j^3}{3(1-\rho_{j,t})}+\frac{\Lambda_{j,t}^2\Gamma_j^4}{4(1- \rho_{j,t})^2} , \label{eq:lemma2_2}
\end{eqnarray}
where $\rho_{j,t}=\Lambda_{j,t} / \mu_j$ is the request intensity at node $j$, and $\Lambda_{j,t}=\sum_{i=1}^r \lambda_{i,t}\pi_{i,j,t}$ is the mean arrival rate at node $j$.
The bound is tight in the sense that there exists a distribution of ${\bf Q}_{j,t}$ such that (\ref{eq:lemma1}) is satisfied with exact equality.
\end{lemma}
{
\begin{proof}
	The proof follows on the same lines as in \cite{Yu_TON}. A brief detail into the proof idea is provided in Appendix \ref{apdx} ( where the time-index $t$ is omitted) to make the paper more self-contained.
	\end{proof}}
{
\begin{remark} The upper bound of Lemma 1 can be improved if higher ordered moment statistics of service time are known, though we do not consider them in this work due to tractability in latency optimization. This bound, even without the additional moments, has been shown to outperform existing bounds and is indeed very close the actual latency  \cite{Yu_TON}.
\end{remark}
}
{Under the assumption of time-scale separation, each time bin has stationary request arrival rates and is long enough for cache content to reach a steady state. We consider the cache optimization in a single time bin to minimize weighted mean latency quantified by Lemma 1. This optimization will be performed repeatedly at the beginning of each time bin to address time-varying request arrival rates during busy and off-peak hours.}

We now formulate the cache optimization in a single time-bin. The optimization is over cache content placement $d_{i,t}$, scheduling probabilities $\pi_{i,j,t}$, and auxiliary variable $z_{i,t}$ in the upper bound. Let $\hat{\lambda}_t=\sum_{i=1}^r \lambda_{i,t}$ be the total arrival rate, so $\lambda_{i,t}/\hat{\lambda}$ is the fraction of file $i$ requests, and average latency of all files is given by $\sum_{i=1}^r (\lambda_{i,t}/\hat{\lambda}_t)\bar{T}_{i,t}$. Our objective is to minimize an average {\em latency} objective, i.e.,
\vspace{-0.1in}
\begin{eqnarray}
& \D {\rm min} & \sum_{i=1}^r \frac{\lambda_{i,t}}{\hat{\lambda}_t}  \bar{U}_{i_t}  \label{eq:JLRM-SC}\\
& {\rm s.t.} &  (\ref{eq:lemma1}), \ (\ref{eq:lemma2_1}), \ (\ref{eq:lemma2_2}), \  \sum_{j=1}^{m}\pi_{i,j,t} = k_i - d_{i,t}, \  \pi_{i,j,t}, d_{i,t}\ge 0, \nonumber\\ && \sum_{i=1}^r d_{i,t}\le C, \  \pi_{i,j,t} = 0 \text{ for } j\notin {\cal S}_i, \  \pi_{i,j,t}\le 1, \nonumber\\
&&z_{i,t}\ge 0,  d_{i,t} \in {\mathbb Z}.  \label{eq:JLRM-SC2}   \nonumber \\
& {\rm var.} &  \pi_{i,j,t}, \ d_{i,t}, \ z_{i,t} \ \forall i,j,t. \nonumber
\end{eqnarray}
Here the constraints  $\sum_{j=1}^{m}\pi_{i,j,t} = k_i - d_{i,t}$ and $\pi_{i,j,t}\le 1$ ensure that  $k_i - d_{i,t}$ distinct storage nodes (along with $d_{i,t}$ chunks in the cache) are selected to process each file request, following probabilistic scheduling in \cite{Yu_TON}. Clearly, storage nodes without desired chunks cannot be selected, i.e., $\pi_{i,j,t} = 0 \text{ for } j\notin {\cal S}_i$. Finally, the cache has a capacity constraint $\sum_{i=1}^r d_{i,t}\le C$.

Solving the cache optimization gives us the optimal cache content placement and scheduling policy to minimize file access latency. We note that the constraint $z_{i,t}\ge 0$ is not needed if none of the files is completely in the cache. However, the latency bound does not hold if the file is completely in the cache since in that case the bound is $z_{i,t}$ in the above expression. In order to avoid having indicators representing the constraint on $z_{i,t}=0$ if the file is in the cache, we only consider $z_{i,t}\ge 0$ making the latency bound hold irrespective of the number of chunks in the cache.  This problem can be rewritten as follows.

\noindent \hspace{0.2in} {\bf Distributed Storage with Caching:}

\begin{eqnarray}
& {\rm min} &  \sum_{i=1}^r \lambda_{i,t}z_{i,t}/\hat{\lambda}_t \nonumber\\&&+ \sum_{i=1}^r \sum_{j=1}^m \frac{\lambda_{i,t}\pi_{i,j,t}}{2\hat{\lambda}_t} \left[ X_{i,j,t} + \sqrt{X_{i,j,t}^2 + Y_{j,t}} \right]  \label{eq:c0} \\
& {s.t.} & X_{i,j,t}=   \frac{1}{\mu_j} + \frac{ \Lambda_{j,t} \Gamma_j^2 }{2(1- \rho_{j,t})}-z_{i,t}, \ \forall j  \label{eq:c1}   \\
&  & Y_{j,t}=  \sigma_j^2+\frac{ \Lambda_{j,t} \hat{\Gamma}_j^3}{3(1-\rho_{j,t})}+\frac{\Lambda_{j,t}^2\Gamma_j^4}{4(1- \rho_{j,t})^2}, \ \forall j \label{eq:c2}  \\
& & \rho_{j,t} = \Lambda_{j,t}/\mu_j < 1, \ \Lambda_{j,t}=\sum_{i=1}^r \pi_{i,j,t}\lambda_{i,t}  \ \forall j \label{eq:c3} \\
&  & \sum_{j=1}^m \pi_{i,j,t} = k_i-d_{i,t}, \ \pi_{i,j,t}\in [0,1], z_{i,t}\ge 0,\\&& \pi_{i,j,t}=0 \ \forall j\notin \mathcal{S}_i, \sum_{i=1}^r d_{i,t}\le C, \  d_{i,t}\in \mathbb{Z}^+ \label{eq:c4}  \\
& {\rm var.} & z_{i,t}, \ d_{i,t}, \ \pi_{i,j,t}, \ \forall i,j. \nonumber
\end{eqnarray}

%% file: Algo.tex
\subsection{Proposed Algorithm}
The proposed cache optimization problem in \eqref{eq:c0}-\eqref{eq:c4} is an integer optimization problem, since the number $d_{i,t}$ of functional chunks in the cache must be integers. To solve this problem, we propose a heuristic algorithm, which iteratively identifies the files that benefit most from caching, and constructs/stores funtional chunks into cache memory accordingly. We first note that the variable $d_{i,t}$ can be absorbed into scheduling decision $\pi_{i,j,t}$ because of the equality constraint $d_{i,t}=k_i-\sum_{j=1}^m \pi_{i,j,t}$. Thus, there are two set of variables -  $z_{i,t}$, and $\pi_{i,j,t}$ - we need to consider. It is easy to show that the objective function is convex in both these variables, however there is an integer constraint on $\sum_{j=1}^m \pi_{i,j,t}$ due to the integer requirement of $d_{i,t}$.

The algorithm employs an alternating minimization over two dimensions - the first decides on $z_{i,t}$ given $\pi_{i,j,t}$, and the second decides on $\pi_{i,j,t}$ given $z_{i,t}$. The first problem is convex, and can be easily solved by gradient descent. However, the second problem has integer constraint. In order to deal with this, we first remove integer constraint to solve the problem. Then, a certain percentage of files whose fractional part of content accessed from the disk is highest are added a part in the disk to make the part in the disk as integers. The optimization over  $\pi_{i,j,t}$ keeps running until $\sum_{j=1}^m \pi_{i,j,t}$ for all files is an integer. In particular, we derive the two sub-problems that need to be solved as follows.

We define the problem {\em Prob\_Z} for given $\pi_{i,j,t}$ as
\begin{eqnarray}
& {\rm min} &  \sum_{i=1}^r \lambda_{i,t}z_{i,t}/\hat{\lambda}_t \nonumber\\&&+ \sum_{i=1}^r \sum_{j=1}^m \frac{\lambda_{i,t}\pi_{i,j,t}}{2\hat{\lambda}_t} \left[ X_{i,j,t} + \sqrt{X_{i,j,t}^2 + Y_{j,t}} \right]   \\
& {s.t.} & \eqref{eq:c1}, \eqref{eq:c2} , \eqref{eq:c3},  z_{i,t}\ge 0\nonumber \\
& {\rm var.} & z_{i,t},  \ \forall i. \nonumber
\end{eqnarray}

We define the problem {\em Prob\_$\Pi$} for given $z_{i,t}, \  k_{U,i,t}, \ k_{L,u,t}$ as
\begin{eqnarray}
& {\rm min} &  \sum_{i=1}^r \lambda_{i,t}z_{i,t}/\hat{\lambda}_t \nonumber\\&&+ \sum_{i=1}^r \sum_{j=1}^m \frac{\lambda_{i,t}\pi_{i,j,t}}{2\hat{\lambda}_t} \left[ X_{i,j,t} + \sqrt{X_{i,j,t}^2 + Y_{j,t}} \right]  \\
& {s.t.} & \eqref{eq:c1} , \eqref{eq:c2},  \eqref{eq:c3}, \ \pi_{i,j,t}\in [0,1],  \nonumber \\
&  & K_{L,i,t}\le \sum_{j=1}^m \pi_{i,j,t} \le k_{U,i,t}, \\&& \pi_{i,j,t}=0 \ \forall j\notin \mathcal{S}_i, \sum_{i=1}^r (k_i - \sum_{j=1}^m \pi_{i,j,t})\le C,    \\
& {\rm var.} & \pi_{i,j,t}, \ \forall i,j. \nonumber
\end{eqnarray}

The problem { Prob\_Z} optimizes over $z_{i,t}$ given $\pi_{i,j,t}$. This problem is convex with only one linear constraint $z_{i,t}\ge 0$. In order to solve this problem, we can use standard gradient descent, with making $z_{i,t}$ as zero if the solution is negative in each iteration. The problem Prob\_$\Pi$ assumes that the number of total chunks of a file $i$ accessed from the disk is between $k_{L,i,t}$ and $k_{U,i,t}$. As we decide the number of chunks in the cache for each file, these two bounds will become equal. This problem is also convex, and can be solved using projected gradient descent. With algorithmic solution to these two sub-problems, the algorithm for Distributed Storage with Caching is given in Algorithm 1.

{
\begin{lemma} The 2 sub-problems, Prob\_Z and Prob\_$\Pi$, are both convex. 
	\end{lemma}
}

\begin{figure}
 \vspace{-8mm}
    \begin{minipage}{.47 \textwidth}
      \begin{algorithm}[H]
        \caption{Our Proposed Algorithm for Distributed Storage with Caching}
        \label{alg:MSC}
        \begin{algorithmic}
        \STATE Initialize $c=0$ and feasible $(z_{i,t}, \pi_{i,j,t} \ \forall i,j)$
\STATE Compute current objective value $B^{(0)}$
\STATE Initialize $c=0$ and feasible $(z_{i,t}, \pi_{i,j,t} \ \forall i,j)$
\STATE Compute current objective value $B^{(0)}$
\STATE \textbf{do}
\STATE \hspace{\algorithmicindent} Solve Convex Problem { Prob\_Z} to get $z_{i,t}$ for given $\pi_{i,j,t}$ for all $i$
\STATE \hspace{\algorithmicindent} Set $k_{L,i,t}=0$, $k_{U,i,t} = k_i$
\STATE \hspace{\algorithmicindent}\textbf{do}
\STATE \hspace{\algorithmicindent} \hspace{\algorithmicindent} Solve Convex Problem Prob\_$\Pi$ to get $\pi_{i,j,t}$ for given $z_{i,t}, \  k_{L,i,t}, \  k_{U,i,t}$ for all $i,j$
\STATE \hspace{\algorithmicindent} \hspace{\algorithmicindent} Let $i_1 = \arg\max$ (fractional part of $\sum_{j=1}^m \pi_{i,j,t}$)
\STATE \hspace{\algorithmicindent} \hspace{\algorithmicindent}$k_{L,i_1,t}= k_{U,i_1,t} = ceil(\sum_{j=1}^m \pi_{i,j,t})$
\STATE \hspace{\algorithmicindent} \textbf{while} $\sum_{i=1}^r frac(\sum_{j=1}^m \pi_{i,j,t})>0$
\STATE \hspace{\algorithmicindent}Compute new objective value $B^{(c+1)}$,  Update  $c=c+1$
\STATE {\bf while} $B^{(c)}-B^{(c-1)}>\epsilon$
        \end{algorithmic}
      \end{algorithm}
    \end{minipage}
    \vspace{-.25in}
  \end{figure}

{
\begin{remark} The proposed algorithm 1 is guaranteed to converge. Since the inner loop that solves convex problem Prob\_$\Pi$ iteratively determines the placement of chunks in cache one-by-one, until the entire cache space is filled up, it runs at most $r$ iterations. The outer loop that solves convex optimization Prob\_Z generates a monotonically decreasing sequence of objective values and is also guaranteed to converge within finite number of iterations.
\end{remark}
}

We note that the inner do-while logic to deal with integer optimization runs at most $r$ times. Since $r$ may be large, rather than choosing one index $i_i$, we choose a ceiling of certain fraction of file indices among those which have fractional content in the cache. This makes the loop run in $O(\log r)$. Thus, each outer loop runs $O(\log r)$ convex problems. The algorithm will be solved repeatedly for each time bin to guide the update of cache content for service latency minimization. 

%% file: Imple2.tex
\section{Numerical Results, Implementation and Evaluation}

In this section, we evaluate our proposed algorithm for functional cache optimization, through both simulation and implementation in an open-source, distributed filesystem.

\subsection{Numerical Results Setup}
We simulated our algorithm in a cluster of $m=12$ storage servers, holding $r=1000$ files of size 100 MB each using a (7,4) erasure code. Unless stated otherwise, cache size remains as 500 times of the chunk size (i.e., 500 times of 25 MB). The arrival rate for each file is set at a default value of $\lambda_i=$ 0.000156/sec,  0.000156/sec, 0.000125/sec, 0.000167/sec, 0.000104/sec for every five out of the 1000 files of each size. It gives an aggregate arrival rate of all files to be 0.1416/sec. The inverse of mean service times  for the 12 servers are set based on measurements of real service time in a distributed storage system, which we obtained from a previous work \cite{Yu_TON}, and they are \{0.1, 0.1, 0.1, 0.0909, 0.0909, 0.0667, 0.0667, 0.0769, 0.0769, 0.0588, 0.0588\} for the 12 storage servers respectively. The placement of files on the servers is chosen at random, unless explicitly specified. The major objective of our simulation is to validate our latency analysis of erasure-coded storage with caching in those areas that are hard to implement in a real test-bed or are difficult to measure in experiments.
\subsection{Numerical Results}

\begin{figure}[!thbp]
\vspace{-2mm}
\begin{center}
{\includegraphics[trim=1.65in 3.3in 1.8in 3.5in, clip, width=.4\textwidth, draft=false]{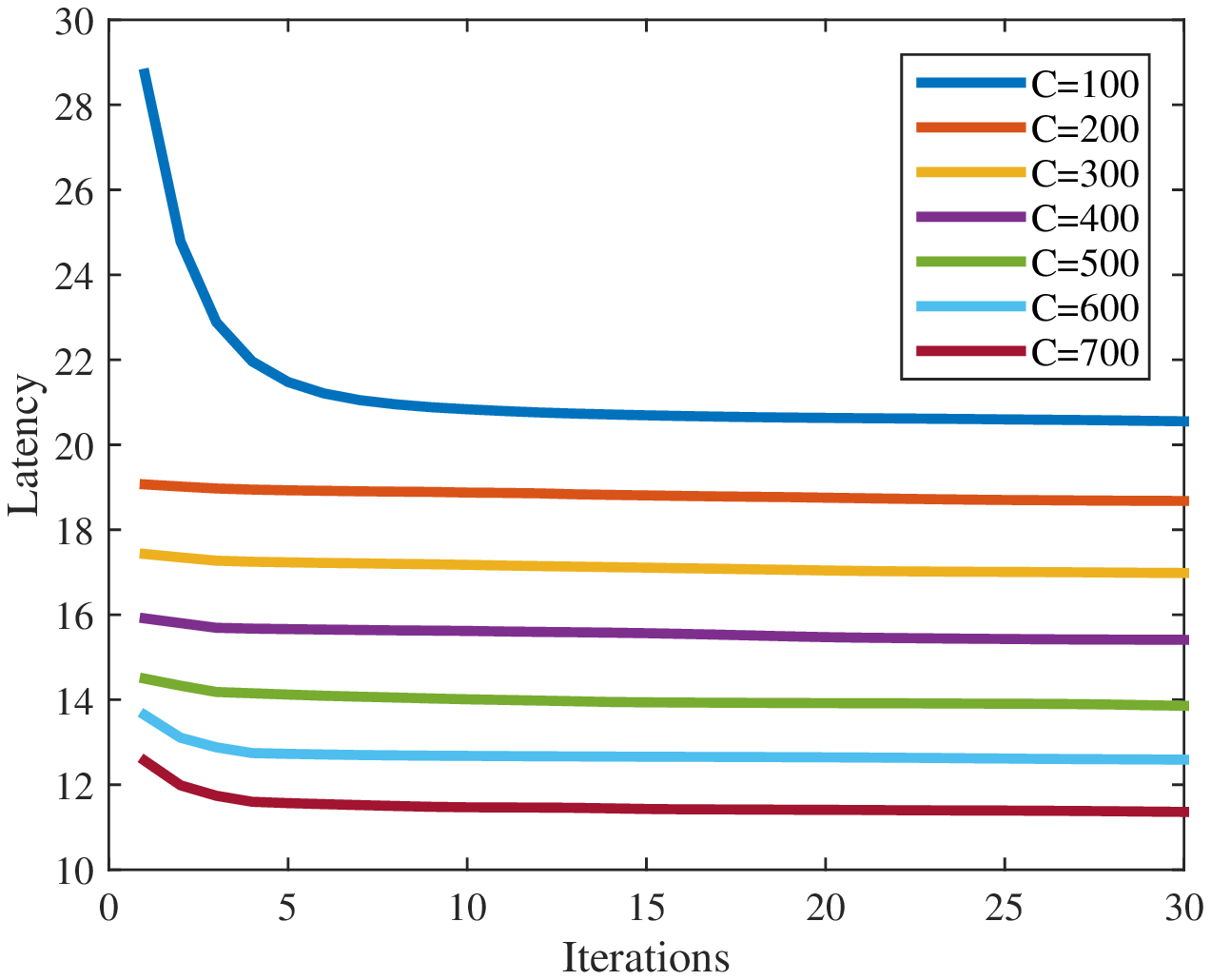}}
\vspace{-3mm}
\caption{Convergence of the proposed algorithm for a system with $r = 1000$ files each of size 100MB and using a cache size of $C\times 25$M. The algorithm efficiently computes a solution in less than 20 iterations with a low tolerance $\epsilon = 0.01$ for different cache sizes.  A random initialization is chosen for C = 100, while the converged solution for C = 100 is taken as initialization for C = 200 and so on.}
\label{fig:conv}
\end{center}
\vspace{-.2in}
\end{figure}

{\bf Convergence of Algorithm:} We implemented our cache optimization algorithm using MOSEK, a commercial optimization solver, to project the gradient descent solution to the feasible space for  Prob\_$\Pi$. For 12 distributed storage servers in our testbed, Figure \ref{fig:conv} demonstrates the convergence of our algorithm in one time-bin, which optimizes the average latency of all files over request scheduling $\pi_{i,j,t}$ and cache placement $d_{i,t}$ for each file. We assume 1000 files, each using a (7,4) erasure code and of size 100 MB, divided evenly into five groups with the arrival rates of the five groups as mentioned above. The convergence of the proposed  algorithm is depicted in Fig. \ref{fig:conv} for cache size $C\times 25$ MB. For $C=4000$, four chunks of each file can be in the cache. A random initialization is chosen for $C=100$, while the converged solution for $C=100$ is taken as initialization for $C=200$ and so on. We note that the algorithm converges within a few iterations, and converges in less than 20 iterations with a threshold of 0.01 on latency for all cache size values in Figure \ref{fig:conv}.


\begin{figure*}[!thp]
\vspace{-3mm}
\begin{minipage}{0.31\textwidth}
\begin{center}

{\includegraphics[trim =1.6in 2.5in 1.8in 3.55in, width = \textwidth, draft=false]{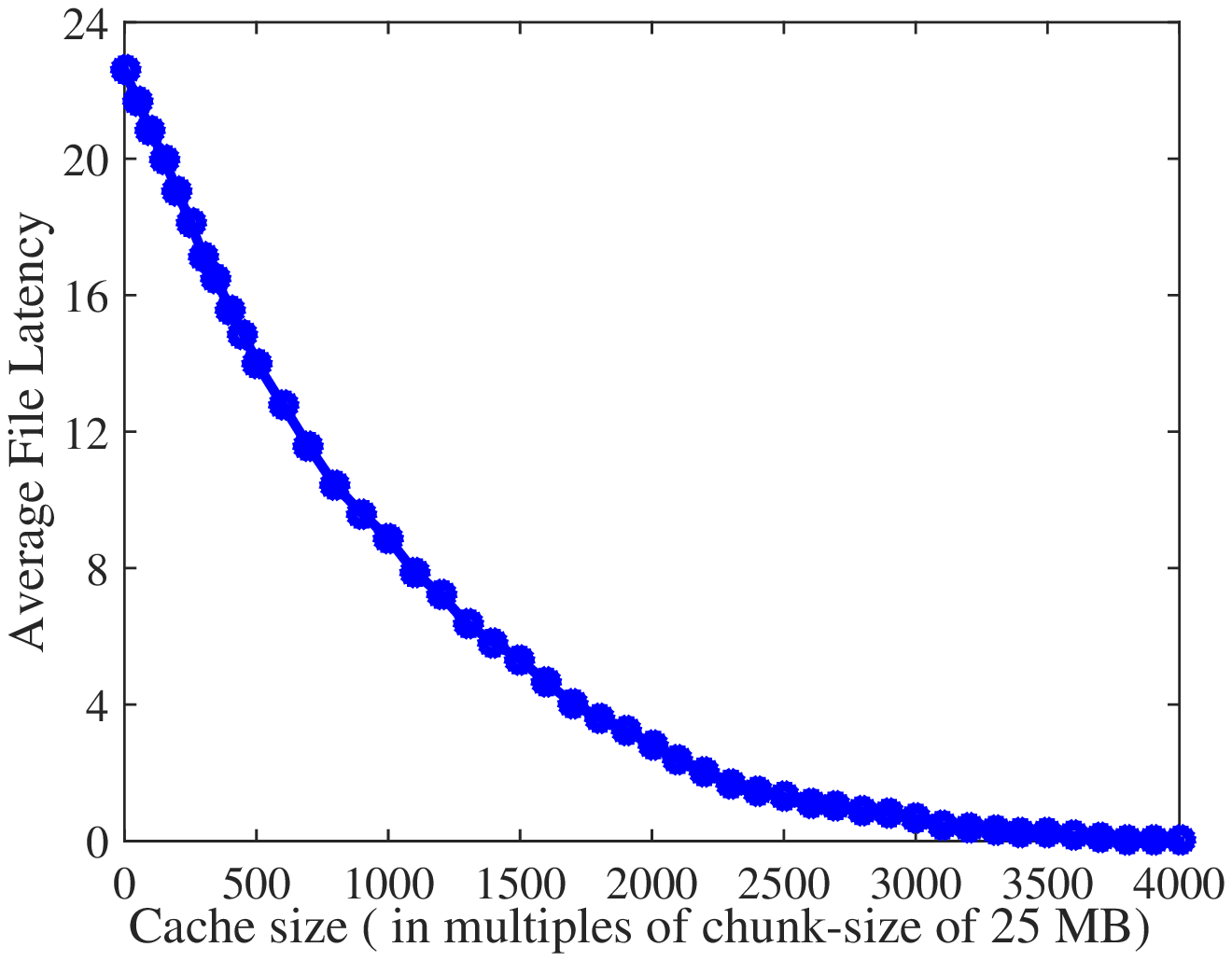}}

\vspace{-13mm}
\caption{Average latency as cache size varies from 0 to 4000 chunk-size, each data point represents average latency for $r=1000$ files in one time bin. This clearly demonstrates that the average latency decreases as cache size increases when all other parameters are fixed. }
\label{fig:cache}
\end{center}
\end{minipage}
\hspace{0.2cm}
\begin{minipage}{0.31\textwidth}
\begin{center}
{\includegraphics[trim =0.3in 1.1in 0.4in 2.4in, clip, width = \textwidth, draft=false]{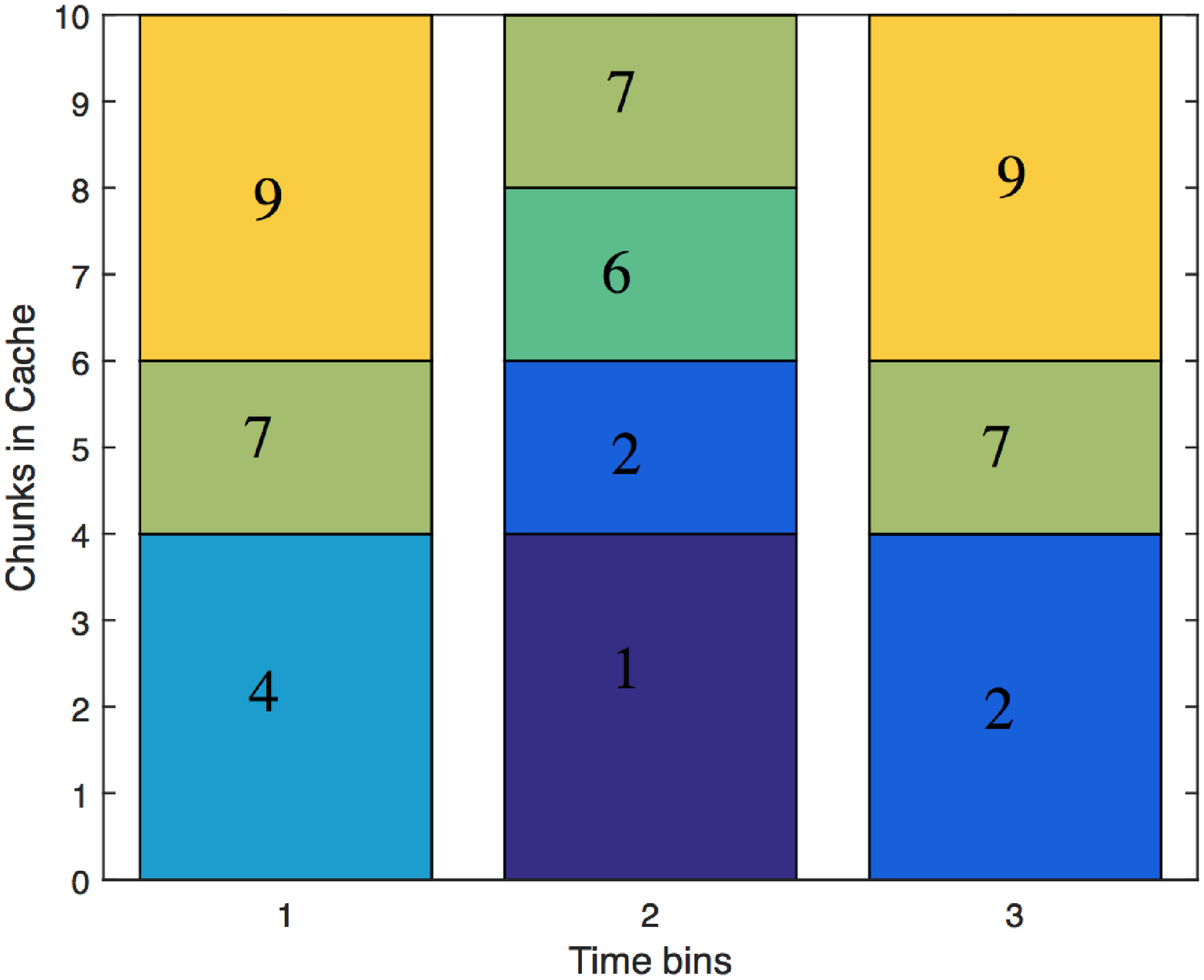}}
\vspace{-17mm}
\caption{Evolution of cache content in three time bins, each having different request arrival rates for the $r=10$ files. It shows that our algorithm is updating content held in the cache according to the intensity of workload of each file effectively. }
\label{fig:cache_change}
\end{center}
\end{minipage}
\hspace{0.2cm}
\begin{minipage}{0.34\textwidth}
\begin{center}
{\includegraphics[trim = 1.7in 2.2in 1.8in 3.5in, clip, width = \textwidth, draft=false]{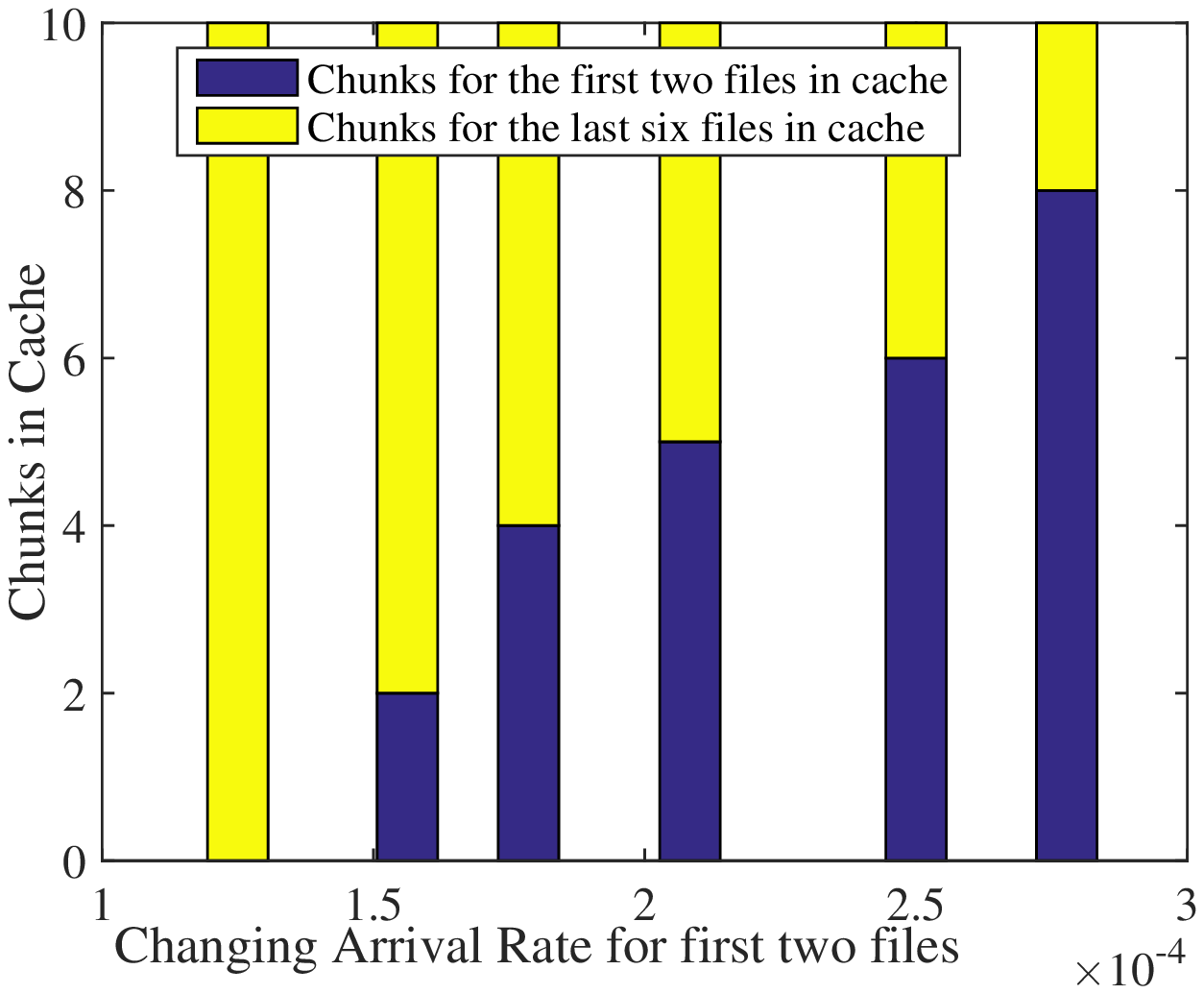}}
\vspace{-20mm}
\caption{The chunk placement depends not only on arrival rates but also on the placement of contents in the storage nodes and the service rates.  }
\label{fig:arrrate_chunk}
\end{center}
\end{minipage}
\vspace{-.2in}
\end{figure*}

{\bf Impact of Cache Size:} To validate that our proposed algorithm can effectively update file content placement in cache to optimize average latency in the storage system, we plot average latency for $r=1000$ files of size 100MB in one time bin with designated request arrival rate for each file, while cache size varies from 0 to 4000 times of the chunk-size. Fig \ref{fig:cache} shows that the average latency decreases as cache size increases, where average latency is 23 sec when no file has any content in the cache, and is 0 sec when the cache size is 4000 chunk-size since 4 chunks of each file can be in the cache. We note that the latency is a convex decreasing function of the cache size, depicting that our algorithm is effectively updating content in the cache and showing diminishing returns in a decrease of latency after reaching certain cache size.

{\bf Evolution of Cache Content Placement:} We validate that our algorithm minimizes service latency by optimizing the cache content with respect to request arrival rates (the algorithm is minimizing latency bound $\bar{U}_{i,t}$, and $\lambda_i$ and $\Lambda_j$ is playing an important role in $\bar{U}_{i,t}$), i.e., placing more file chunks that are frequently accessed (with a higher arrival rate) in the cache. We design a simulation which consists of 10 files, of size 100MB, and run our algorithm for three time bins, where each time bin has a different request arrival rate for the $r=10$ files as given in Table \ref{tab:arr}. { The arrival rate changes are detected through a sliding-window-based mechanism, which continuously measures the average request arrival rates, and when the difference of arrival rates of an object between two different time windows is greater than a threshold, a new time bin is introduced to trigger a new round of optimization.} The arrows in Table \ref{tab:arr} indicate the increase or decrease of the file arrival rate in the consecutive time bins.
\begin{table*}
\caption{Table of arrival rates for 10 files in 3 time bins.} \label{tab:arr}
\centering
    \begin{tabular}{|c|c|c|c|c|c|c|c|c|c|c|}
    \hline
    Time Bin & File 1 & File 2 & File 3 & File 4 & File 5 & File 6 & File 7 & File 8 & File 9 & File 10 \\ \hline
    1 & 0.000156	& 0.000156 & 0.000125 & 0.000167 & 0.000104 & 0.000156 & 0.000156 & 0.000125 & 0.000167 & 0.000104 \\ \hline
    2 & 0.000156	& 0.000156 & 0.000125 & 0.000125 \color{red}{$\downarrow$} & 0.000125 {$\uparrow$} & 0.000156 & 0.000156 & 0.000125 & 0.000125 \color{red}{$\downarrow$} & 0.000125 {$\uparrow$} \\ \hline
    3 & 0.000125 \color{red}{$\downarrow$} & 0.00025 {$\uparrow$} & 0.000125 & 0.000167 & 0.000104 & 0.000125 \color{red}{$\downarrow$} & 0.00025 {$\uparrow$} & 0.000125 & 0.000167 & 0.000104 \\
    \hline
    \end{tabular}
    \vspace{-.25in}
\end{table*}
In this simulation, we plot the cache placement for the 10 files in steady state (after the algorithm converges) for each time bin. From Figure \ref{fig:cache_change}, we see that in the first time bin, as file 4 and file 9 had the highest request arrival rates, they had the highest numbers of chunks in the cache in the steady state. In the second time bin, as  the arrival rates of files 4 and 9 decreased, and  those of files 5 and 10 increased, now the files that had the same and higher request arrival rate were 1, 2, 6, and 7. Thus Fig \ref{fig:cache_change} shows that in the second time bin these four files dominates the cache's steady state. In the third time bin, the files that had the highest arrival rate became files 2, 7, 4, and 9, and Fig. \ref{fig:cache_change} also shows that cache is mainly holding the contents of file 2, 7, and 9. Due to different service rates of the servers and the randomized placement of contents in the servers, it is not always the case that files with the highest arrival rate need to be placed completely in the cache. Rather, it is important to determine how many chunks of each file should be placed in the cache.


{\bf Impact of Content Placement and Arrival Rate:} We note that in addition to the arrival rates of the files, placement of content on the storage nodes influence the decision of cache placement. We consider 10 files, of size 100M, using (7,4) erasure code are placed on 12 servers as described in the simulation set up. The first three files are placed on the first seven servers while the rest of the files are placed on the last seven servers. Note that servers 6 and 7 host chunks for all files. For the arrival rates of the files, we fix the arrival rate for the last eight files such that the arrival rate of the third and fourth file is $.0000962$/sec, and of the last six files is 0.0001042/sec. The arrival rates for the first two files are assumed to be the same, and due to symmetry of files and placement, we consider four categories of content placement in the cache - contents for the first two files, contents for 3rd file, contents for 4th file, and the contents for the last six files. In the considered arrival rates in Figure \ref{fig:arrrate_chunk}, there was no chunk for the third and the fourth file placed in the cache due to low arrival rates. We note that we always assume the arrival rates of the first two files as the highest but since the servers on which they are placed have relatively lower average load, the arrival rate needs to be significantly higher for them to increase content in the cache. The six bars in Figure \ref{fig:arrrate_chunk} correspond to arrival rates for the first two files of $0.0001250,    0.0001563,    0.0001786,    0.0002083,    0.0002500,$ and     0.0002778, respectively. At an arrival rate of $.000125$/sec for the first two files, there is no content for these files in the cache. Thus even though the arrival rate for the first two files is the largest, they do not need to be placed in the cache due to the fact that they are placed in the storage servers that are lightly loaded.  As the arrival rate increases to $.00015625$/sec, two chunks corresponding to these files start to appear in the cache. Further increase of the arrival rate leads to more and more chunks in the cache. Thus, we note that the placement of chunks on the storage nodes, arrival rates, and service rates all play a role in allocating chunks of different files in the cache. 

\begin{figure}[!thbp]
\begin{center}
\scalebox{0.34}{\includegraphics[draft=false]{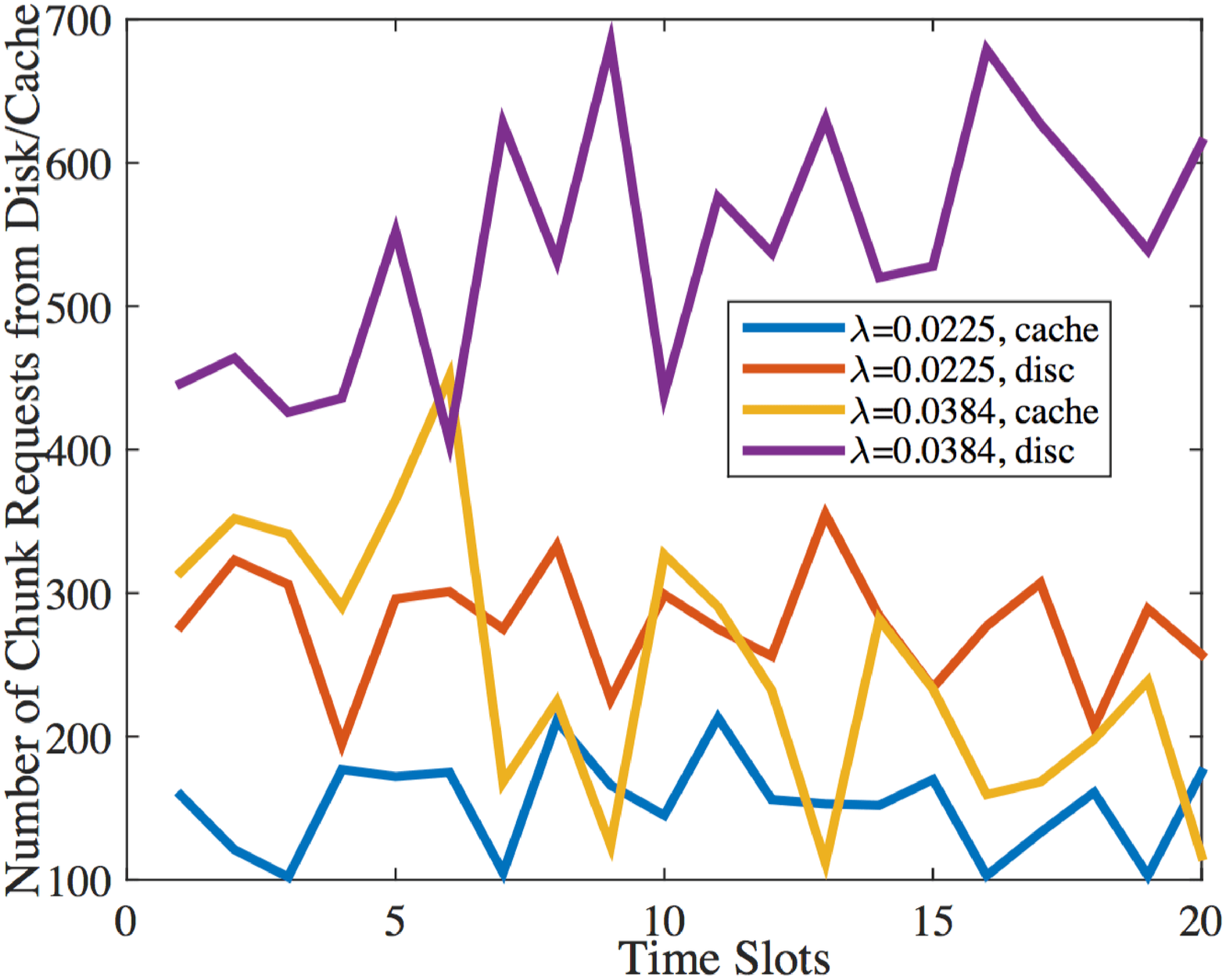}}
\caption{Number of chunk requests that are sent to storage servers or cache with an average arrival rate $\lambda_i=0.0225$ /s and $\lambda_i=0.0384$, respectively, where a time bin of length 100 s is divided into 20 equal time slots. The figure shows the dynamic nature of the number of chunks the client obtained from the cache/nodes. }
\label{fig:dynam}
\end{center}
\end{figure}

{\bf Chunk Request Scheduling Evolution:} In order to see how the request scheduling evolves during each time bin, we run the experiment with $r=1000$ objects, each of size 200 MB and using a (7,4) erasure code, and a total cache size of 62.5 GB. The average file request arrival rate for the two experiments  is chosen as $\lambda_i=0.0225$ /sec and $\lambda_i=0.0384$, respectively. We divide each time bin (100 sec) into 20 time slots, each with a length of 5 sec, and plot the number of chunks the client is getting from the cache and from the storage nodes in each time slot. Fig \ref{fig:dynam} shows that under both workloads, the number of chunks retrieved from the cache is smaller than that from OSDs. As the cache size is 62.5 GB, with chunk size 50 MB, it has a capacity of 1250 chunks, which means each file has more chunks in OSDs than that in cache on an average. Further, since the arrival rate of all file increases proportionally, the relative percentage of chunks retrieved from cache over 100s stays almost the same in the two cases, at about 33\%.

{\subsection{Ceph Testbed} 
We have validated the effectiveness of our latency analysis and the algorithm for joint optimization on average latency in erasure-coded storage systems with caching through simulation. In the rest of this section, we will evaluate the performance of our solution by implementing our algorithm in {\em Ceph} (the Jewel version with filestore), which is an open-source, object-based distributed storage system \cite{Weil}.  Details on Ceph storage API's and cache tier are given in Appendix \ref{testbed}.

{\bf Our Solution:} Our algorithm requires a caching tier that can store erasure-coded chunks in cache based on the proposed optimization. Since Ceph cache tier only provides replicated caching capability, we leverage it to implement an equivalent version of our functional caching policy. 
Typically we use a faster device (SSD) for cache, and by comparing the read latency from an OSD backed with HDD (shown in Table \ref{tab:servicetime}) and SSD (shown in Table \ref{tab:ssd}), we can see that service latency from cache can be negligible compared to that from back-end storage pool. As defined in Equation \eqref{eq:T_bar}, the read latency of a file in erasure coded storage depends on the largest retrieval delay of the chunks (this is also true in Ceph). So the file latency depends only on the retrieval latency of chunks stored in back-end storage. Then the read process of file with an $(n,k)$ code and $d$ chunks in cache becomes equivalent to reading a file with $(n,k-d)$ code (with the same chunk size) from back-end storage pool, while the latency of the $d$ chunks in cache is ignored. 


This allows us to implement our functional caching policy using different $(n,k-d)$ codes and measure the resulting latency. More specifically, as the algorithm provides number of chunks of each file that are placed in the cache, we can dynamically adjust the equivalent erasure code in Ceph, by creating pools with the new equivalent erasure code and forward the file requests to the new pools. In this evaluation with Ceph, to be coherent with simulation setup, we create a Ceph cluster with 12 OSDs, using $(7,4)$ original erasure code with $r=1000$ files, which means as we put more chunks in cache, there will be 5 pools in total: $(7,4-d)$ for $d=0,\ldots,4$. And all of the five pools are backed by the same set of OSDs in the Ceph cluster. So that this environment provides 12 OSDs for object storage, each of the 1000 files has 7 chunks in the storage pool, and 0 to 4 chunks in the cache (changes with workload/timebin). So when a client tries to access a file, the file will be accessed from one of the five pools according to its equivalent code (calculated from the number of chunks in the cache of each file), from the algorithm at that time. 

\subsection{Experiment Setup}
{\bf Testbed Configuration: } Throughout the evaluation, we compare the performance of our algorithm for optimal caching with a baseline: Ceph's caching tier, replicated caching with LRU. We have a test bed with 3 nodes for the Ceph object storage cluster, and another four nodes each with one virtual machine (VM) as  4 Ceph clients, each connected with 10G Ethernet network interface. All OSDs in this testbed is formatted with an ext4 filesystem. The setup for the two cases can be described as follows: (also shown in Fig \ref{fig:testbed})

\noindent 1.  {\em Optimal Erasure-Coded Caching:} Each node has 4 OSDs, and each OSD is backed with a HDD drive of size 900G. There are 5 pools created with the same set of 12 OSDs. Simulated cache capacity of 2500 chunks of 16 MB files.(1000 files has 4000 chunks in total, cache capacity 10G). Each node has one SSD drive that provides four 5G partitions for the journal of each OSD. The number of placement group for each pool is calculated from Equation (\ref{eq:npg}), which is 256.

\noindent 2.  {\em Ceph Replicated Caching with LRU:} Cache tier consists of 2 OSDs, each backed with a 5G SSD partition, cache mode write-back with a capacity of 10G. Cache tier uses dual replication (close to the redundancy provided by (7,4) erasure code). Storage tier has the same set as that in optimal erasure-coded caching. There is one pool created with (7,4) erasure code on the 12 OSDs. The SSD partitions come from different SSD drives on one node. The number of placement group in cache tier is 128 from Equation (\ref{eq:npg}). Journal settings and number of placement groups in the storage pool are the same with the optimal caching case. 

\begin{figure}[!thbp]
\vspace{-2mm}
\begin{center}
\scalebox{0.29}{\includegraphics[draft=false]{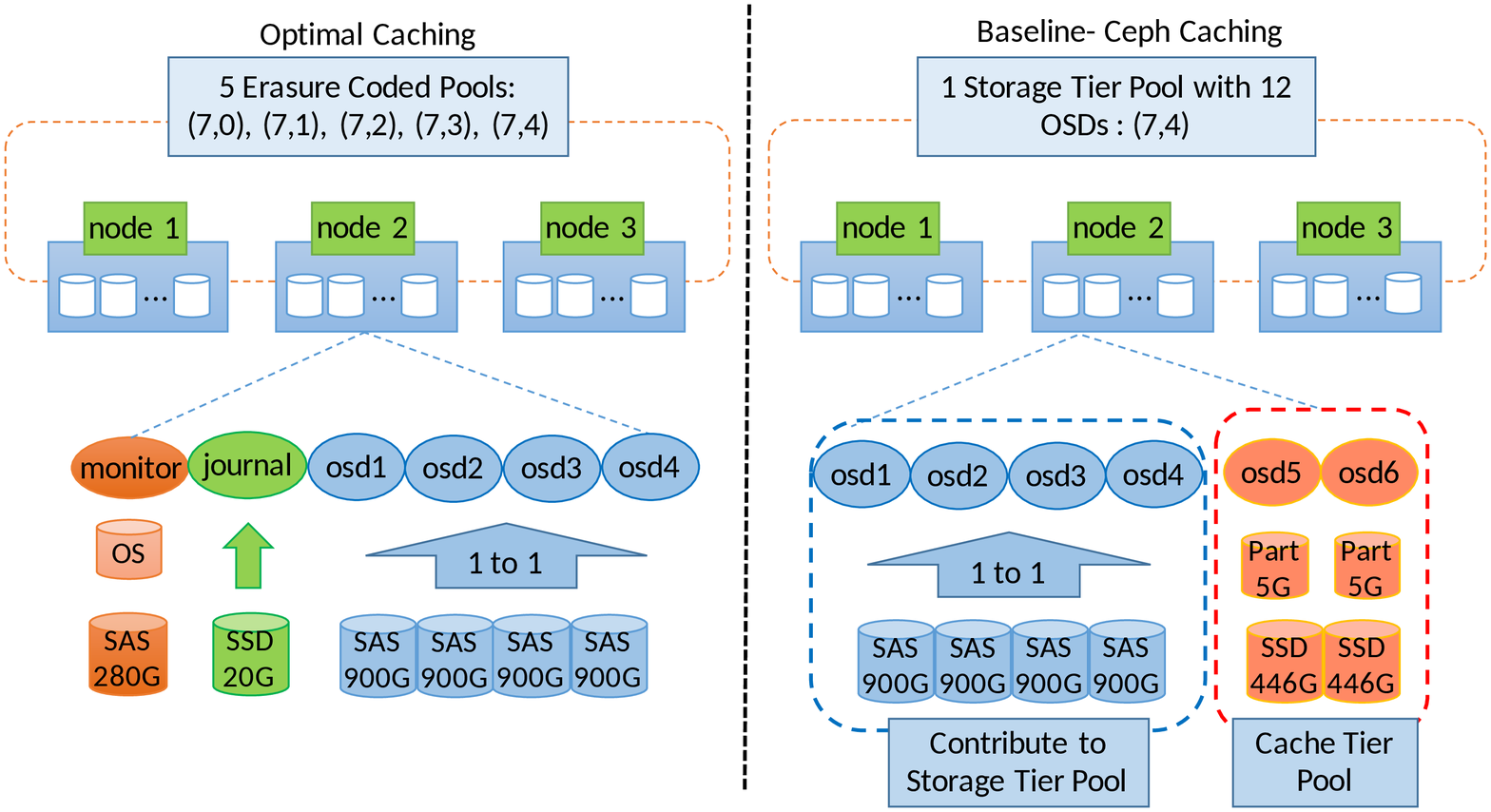}}
\vspace{-2mm}
\caption{Our Ceph Testbed with 3 nodes for proposed optimal caching and ceph's LRU caching as a baseline.}
\label{fig:testbed}
\end{center}
\end{figure}

{\bf Benchmarking Our Ceph Testbed:} We use COSBench 0.4.2 (\cite{bench}, a benchmarking tool for cloud object storage developed by Intel) to benchmark the latency performance (response time in COSBench) in this evaluation. As file requests can be forward to any of the five storage pools according to the algorithm dynamically, to ensure that a file can be accessed from any pools at any time, we perform a write bench mark without cleaning-up to the pools before we run any read benchmarks. We modified COSBench to use existing pools instead of creating new pools (which is by default). We started 4 COSBench drives on the 4 VM clients.  Each COSBench workload has 5 stages: Initial, Prepare, Main, Cleanup, and Dispose. For obtaining latency performance, the number of threads (workers in COSBench) in each driver in the first three stages is set to 1. The list of parameters is shown in Table \ref{tab:para}. 
\begin{table}[th!]
\caption{Table of COSBench Default Configuration} \label{tab:para}
\vspace{-3mm}
\begin{center}
\begin{tabular}{|c|c|}
\hline
Number of Containers(pools) & 5 or 1 \\ \hline
Number of VM Clients & 4 \\ \hline
Object Size & 4MB to 1GB \\ \hline
Access Pattern & Write, Read \\ \hline
Number of Workers & 1 \\ \hline
Rampup & 30 s \\ \hline
Run time & 1800 s \\     
\hline
\end{tabular}
\end{center}
\end{table}

{ We use COSBench workload generator to emulate a real distributed storage workload, which is synthesized from actual production traces of a large-scale object storage system deployed in a commercial service provider.} We studied the 24-hour object storage workload and choose the most popular object sizes (other similar object sizes are round up or down to the sizes in the table) for our test, and the average request arrival rate for each object in each size during the 24 hours is shown in Table \ref{tab:workload}. 

\begin{table}[th!]
\caption{Table of 24-hour Real Storage Workload.} \label{tab:workload}
\vspace{-3mm}
\begin{center}
\begin{tabular}{|c|c|c|}
\hline
Object Size & Average Request Arrival Rate \\ \hline
4MB & 0.00029868 \\ \hline
16MB & 0.00010824 \\ \hline
64MB & 0.00051852 \\ \hline
256MB & 0.0000078  \\ \hline
1GB & 0.0000024 \\    
\hline
\end{tabular}
\end{center}    
\end{table}

\subsection{Evaluation} 
Now we can evaluate the latency performance of our proposed optimal caching with different types of workloads, compared to Ceph's LRU caching as a baseline.

{\bf Service Time Distribution:} In a Ceph erasure coded pool, a primary OSD receives all write operations, it encodes the object into $n=k+m$ chunks and sends them to other OSDs (according to the value of $n$).  When an object is read from the erasure coded pool, it first checks the status of all OSDs that has a chunk of the required object, and then try to contact all such OSDs for retrieve a chunk. The decoding function can be called as soon as $k$ chunks are read. While our service delay bound applies to arbitrary distribution and works for systems hosting any number of objects, we first run an experiment to understand actual service time distribution on our testbed, which is used as an input in the algorithm. To find the chunk service time distribution for different chunk sizes,  we run a 100\% write benchmark for all object sizes in Table \ref{tab:workload}, into the (7,4) erasure coded pool in baseline testbed ( Ceph Replicated Caching with LRU) without cleaning-up for 900 seconds. Then, we run a 100\% read benchmark for each object size with an average request arrival rate per object (among  all objects of each size) as shown in \ref{tab:workload}. For the five set experiments with five different object sizes, we run each set for 1800 seconds to get enough number of requests to take an average according to arrival rates in Table \ref{tab:workload}. The chunk sizes are 1 MB, 4 MB, 16 MB, 64 MB, 256 MB, accordingly.  We collect the service time of the chunks of different sizes at the OSDs and plot the results. Figure \ref{fig:cdf} depicts the Cumulative Distribution Function (CDF) of the chunk service times for read operations with different chunk sizes. We ignored the result for 1G object (256 MB chunk size) as in the figure as it is much larger than others and it would be hard to see the distribution of the first 4 if we add the result. The average service time at the OSD for read operations with chunk size 256 MB is 6758.06 milliseconds.  Using the measured results, we get the mean and variance of service time of each chunk size as shown in Table \ref{tab:servicetime}. 
\begin{table}[th!]
\caption {Mean and Variance of Service Time for Different Chunk Sizes in MilliSeconds.} \label{tab:servicetime}
\vspace{-3mm}
\begin{center}
\begin{tabular}{|c|c|c|}
\hline
Chunk Size & Mean Service Time & Variance of Service Time \\ \hline
1MB & 6.6696 & 0.0963\\ \hline
4MB & 35.8800 & 2.6925 \\ \hline
16MB & 147.8462 & 388.9872 \\ \hline
64MB & 355.0800 & 1256.6100\\ \hline
256MB & 6758.06 & 554180\\    
\hline
\end{tabular}
\end{center}    
\end{table}
{ We use the obtained mean, and variance of the service time of chunks with different sizes to calculate  moments of the service time distribution that are used to find cache placement in our algorithm. }

\begin{figure}[!thbp]
\vspace{-2mm}
\begin{center}
\scalebox{0.67}{\includegraphics[draft=false]{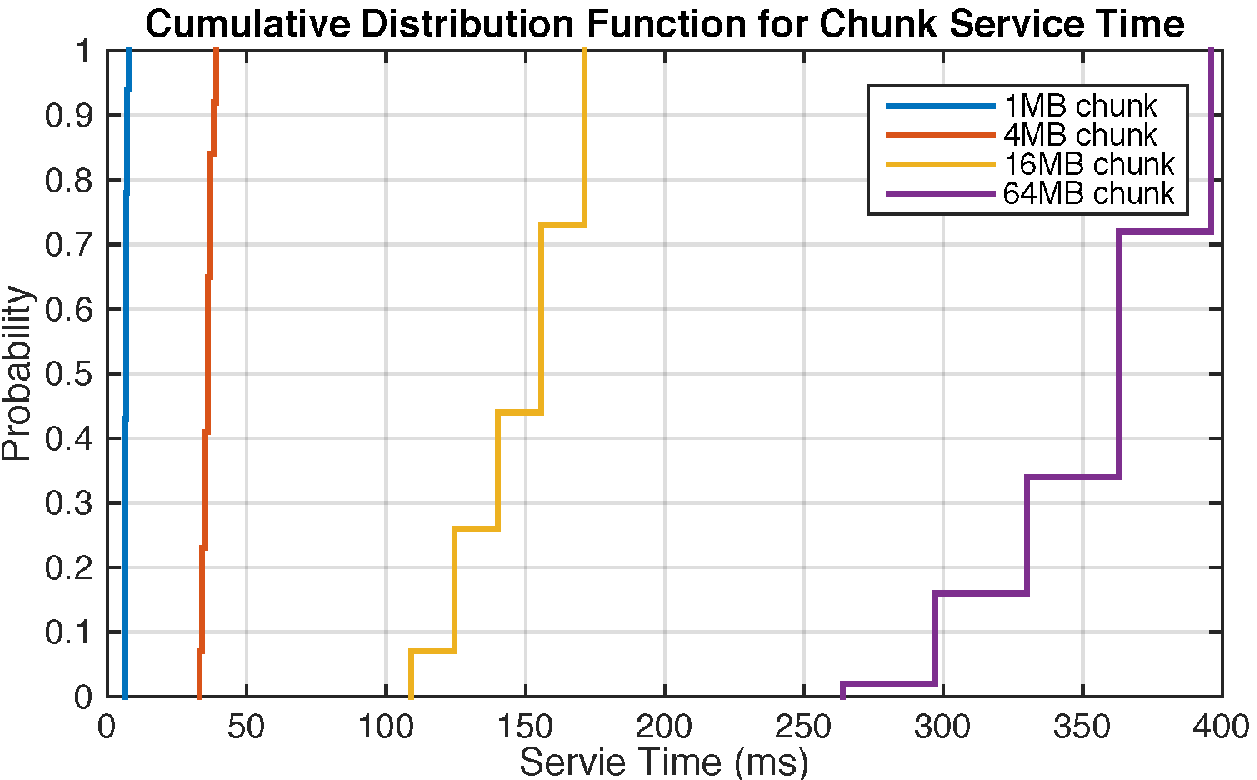}}
\vspace{-3mm}
\caption{Actual service latency distribution for different chunk sizes from the Ceph testbed with 12 OSDs without cache tier.}
\label{fig:cdf}
\end{center}
\vspace{-.1in}
\end{figure}

{\bf Evaluation of the Performance of Our Solution:} As we have validated the algorithm in aspects of convergence and evolution of dynamic caching, we now evaluate the effectiveness of the proposed algorithm (in terms of overall latency improvement) in real storage systems with the real workload. For experiment in optimal caching, as we introduced earlier, we have five pools set up with erasure code  (7,0), (7,1), (7,2), (7,3), (7,4), where pool (7,0) means all the four required  chunks are stored in cache, which will typically be some local SAS or NVMe SSDs in the case of optimal caching. We run a  read test  to the SSD drives for all object sizes in Table \ref{tab:workload} and get the average latency numbers for each object size shown in Table \ref{tab:ssd}. As compared with average service time of the same chunk size from an OSD (HDD as backend) shown in Table \ref{tab:servicetime}, we can see that the latency of a chunk from cache can be negligible when  the read latency of the whole object is measured. This is because the latter depends on the chunk read latency from an OSD in the Ceph cluster, which motivates the use of equivalent codes in the caching schemes. 
\begin{table}[th!]
\caption {Read Latency of different chunk sizes from Cache (SAS SSD) in optimal caching (ms).} \label{tab:ssd}
\vspace{-3mm}
\begin{center}
\begin{tabular}{|c|c|}
\hline
Chunk Size & Read Latency From Cache \\ \hline
1 MB & 1.86619 \\ \hline
4 MB & 7.35639 \\ \hline
16 MB & 30.4927 \\ \hline
64 MB &  97.0968 \\ \hline
256 MB & 349.133 \\ 
\hline
\end{tabular}
\end{center}    
\end{table}

In each time bin, given the workload and the total number of active objects (for simplicity of the algorithm, we set this number to 1000, which is also very close to the number of objects that have been accessed during workload run time for 30-min in the real workload ) in the Ceph object storage cluster, { the optimization algorithm provides which object belongs to which pool (an object-pool map), and the read request arrival rate of each object in each pool. In order to evaluate the effectiveness of the optimal caching algorithm with various object sizes,} first, we run a 100\% write benchmark into the erasure coded pools according to the object-pool map from the optimization for each object size in Table \ref{tab:workload}. Then, for each object size, we run read benchmarks to the erasure-code pools (according to the object-pool map from the optimization) for 1800 seconds. The average read request arrival rate per object in each pool can be calculated as the number of objects in the pool from the algorithm times the average read request arrival rate for each object size in Table \ref{tab:workload}. After the experiment is done, we get the average read latency for objects in each erasure coded pool,  and then calculate the average latency over the 1800 seconds time bin as,
\begin{equation}
\sum_{i} \frac{\text{number of objects in pool } i * \text{average latency in pool } i}{\text{total number of objects in all pools}}.
\label{x}
\end{equation}


For evaluation of Ceph's LRU replicated caching as a baseline, first, we run 100\% write benchmark (with the same number of objects we used in optimal caching) for each object size in Table \ref{tab:workload} to the (7,4) erasure-code storage pool with a replicated cache tier overlay.  Then for each object size, we perform read workloads to this pool with an average read request arrival rate per object as shown in Table \ref{tab:workload} for 1800 seconds. Aggregated request arrival rate in this test for the (7,4) pool would be the request arrival rate in Table \ref{tab:workload} times the total number of active objects (1000 in this case).   In both the cases, cache capacity is fixed and set to be 10GB. This cache can hold 10000 chunks of 4 MB objects, 2500 chunks of 16 MB objects,  625 chunks of 64 MB objects, 156 chunks of 256 MB objects, and 39 chunks of 1 GB objects when using (7,4) erasure code. For example, each object of size 16 MB can have 2500/1000=2.5 chunks on average stored in the cache in both optimal caching and Ceph cache-tier. However, as each object of each size has a different read request arrival rate (value shown in \ref{tab:workload} is an average of all files of the same size), placing two or three chunks in the cache would help reduce latency significantly, thus leading to uneven placement of chunks of the objects in the cache. For this experiment with baseline, we get the average read latency for objects in (7,4) erasure coded pool. We compare the average latency to retrieve an object in optimal caching and Ceph's LRU caching. Fig \ref{fig:file_size} shows that average latency increases as the object size increases, which we can see from our latency bound since the number of chunks that the cache can hold decreases as object size increases. The figure also shows that our caching improves latency as compared to Ceph's LRU caching as a baseline, which is using dual replication in its cache tier, by 26\% on average. This improvement becomes more and more significant as the number of objects increase, which shows that our dynamic optimal chunk placement in the cache is more effective than traditional LRU cache. { The improvement with increased file size is because the load on the system increases thus obtaining more latency advantages with caching. For a lightly-loaded system, the latency is small and the improvement in caching may not be that significant. } Fig. \ref{fig:file_size} shows that our algorithm with optimal caching significantly improves latency with a reasonable number of objects and cache sizes. { Fig. \ref{fig:file_size} also compares the analytically optimized expression of latency from the simulations and the experimentally observed latency where we see that the  analytical result is an upper bound and matches well with the experimental results. }

\begin{figure*}[!th]
\begin{minipage}{0.46\textwidth}
\begin{center}
{\includegraphics[width=\textwidth,draft=false]{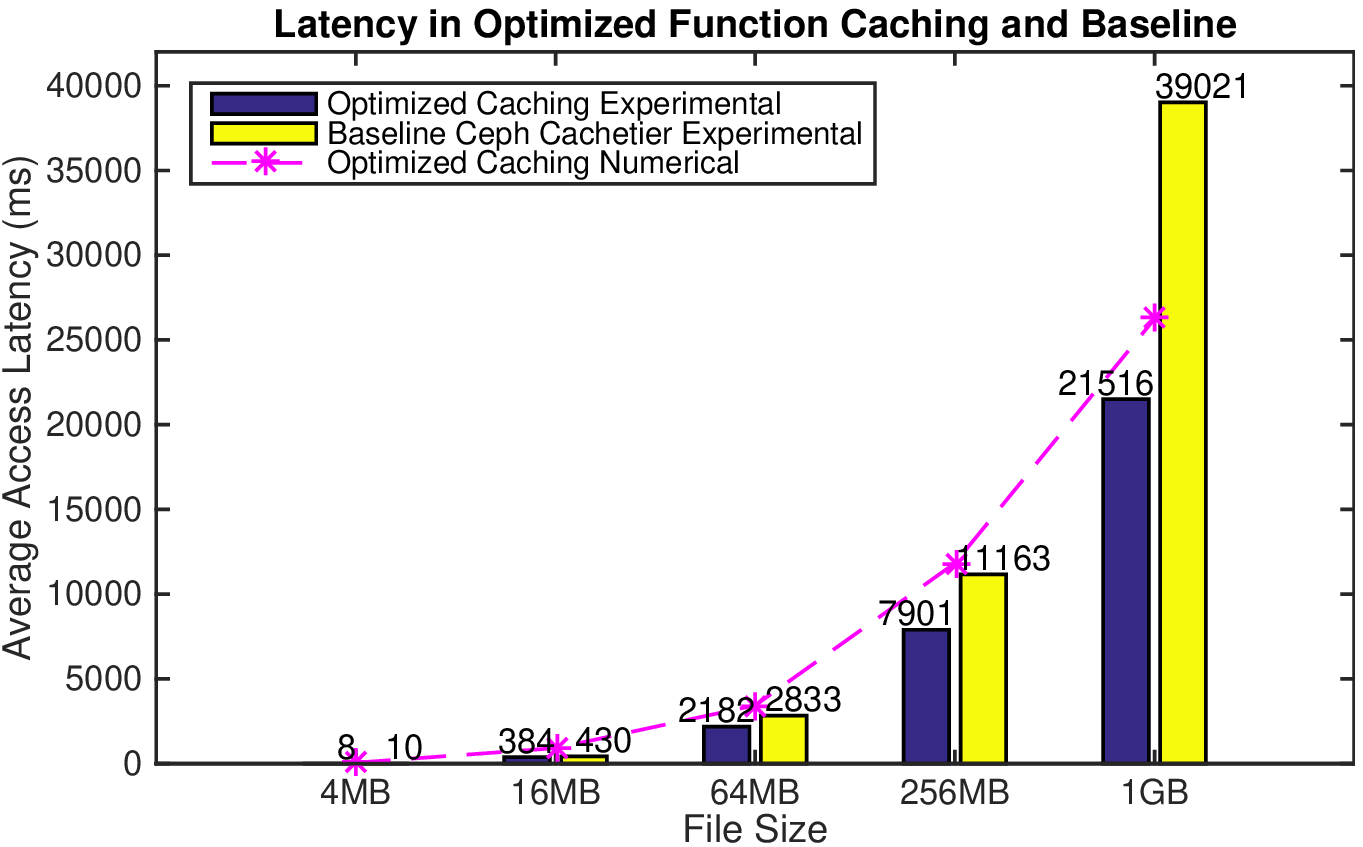}}
\caption{Average latency in optimal erasure coded caching for different object sizes during 1800 seconds benchmark run time, compared with Ceph's LRU replicated caching as a baseline. Cache capacity fixed at 10 GB in both cases. Figure shows significant latency improvement in optimal caching.}
\label{fig:file_size}
\end{center}
\end{minipage}
\hspace{1.2cm}
\begin{minipage}{0.46\textwidth}
\begin{center}
{\includegraphics[width=\textwidth,draft=false]{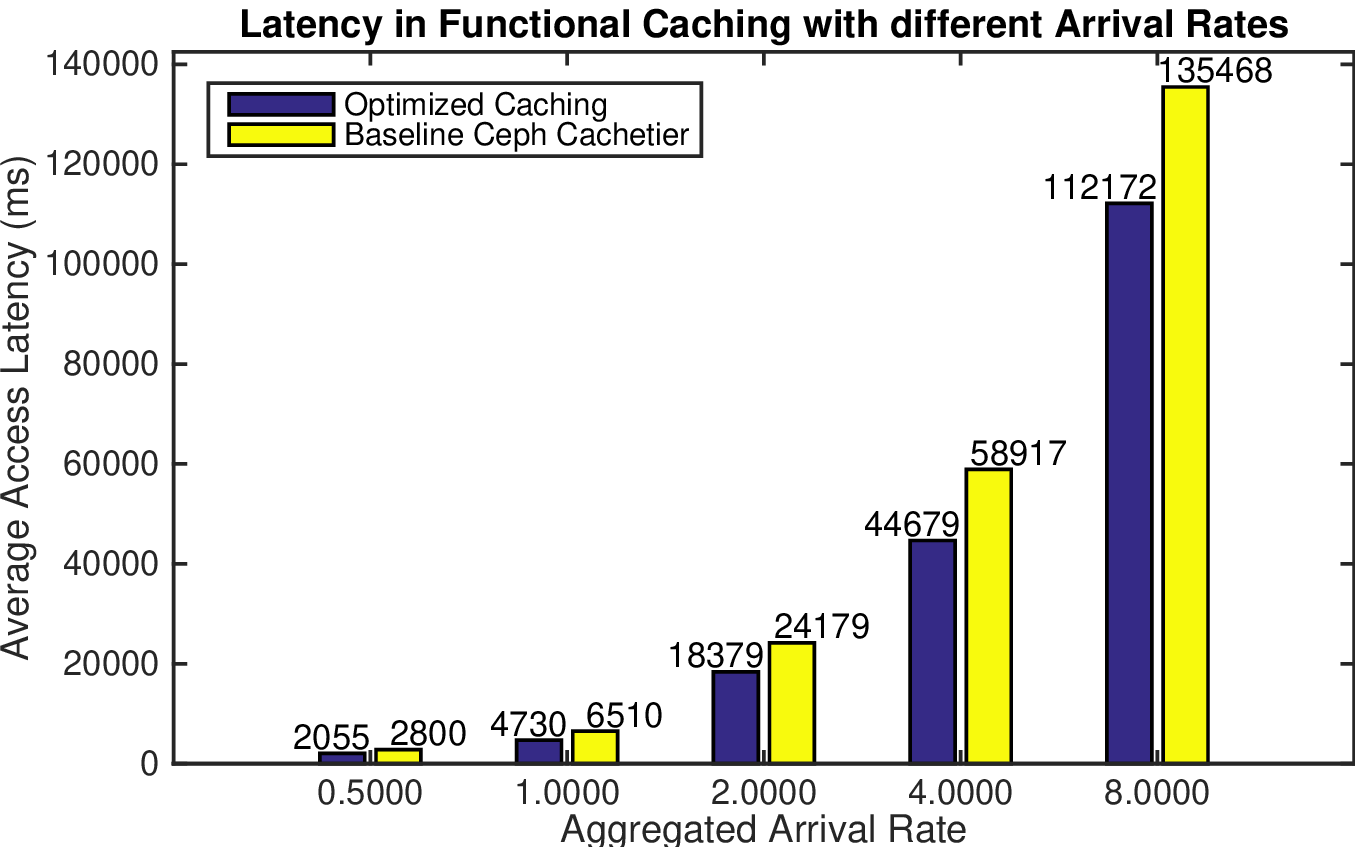}}
\caption{Comparison of average latency in optimal erasure coded caching and Ceph's LRU replicated caching for different intensity of read workload during 1800 seconds benchmark run time. Cache capacity fixed at 10 GB in both cases. Figure shows significant latency improvement in optimal caching.}
\label{fig:workload}
\end{center}
\end{minipage}
\end{figure*}

Next, we evaluate the performance of optimal caching with various workload intensities. We fix the object size at 64 MB (we still use 1000 objects ) and vary the read request arrival rate for these objects and evaluate the latency performance. Queuing latency would not be a dominant issue if the workload is not intense. Thus,  
we fully load the Ceph cluster with much higher request arrival rates as compared to those in Table \ref{tab:workload}. The aggregate read request arrival rates (Single object request arrival rate times the number of objects) used in this test are 0.5, 1.0, 2.0, 4.0, and 8.0. In the optimal caching case, for each workload (request arrival rate), we perform the read benchmark to the five erasure coded pools according to the object-pool map from the optimization with this request arrival rate for 1800 seconds. 
For Ceph's LRU caching, we perform read benchmarks to the (7,4) erasure coded pool with the same set of arrival rates in the case of optimal caching for 1800 seconds, and obtain the average access latency per object. The cache size for both cases is still fixed at 10 GB. Actual average service latency of objects for each workload is shown by a bar plot in Figure~\ref{fig:workload}. In this experiment we also compare the results for the optimal caching scheme and Ceph's LRU caching as a baseline. Fig \ref{fig:workload} shows that our optimal caching algorithm outperforms Ceph's LRU caching in terms of average latency for all workload intensities in the setup. The proposed algorithm gives an average 23.86\%  reduction in latency. Thus, our algorithm with optimal caching can mitigate traffic contention and reduce latency very efficiently compared to Ceph's LRU caching, this also provides a guideline for designing caching schemes under very heavy workloads.}

%% file: bio.tex
\begin{IEEEbiography}[{\includegraphics[width=1in,height=1.25in,clip,keepaspectratio]{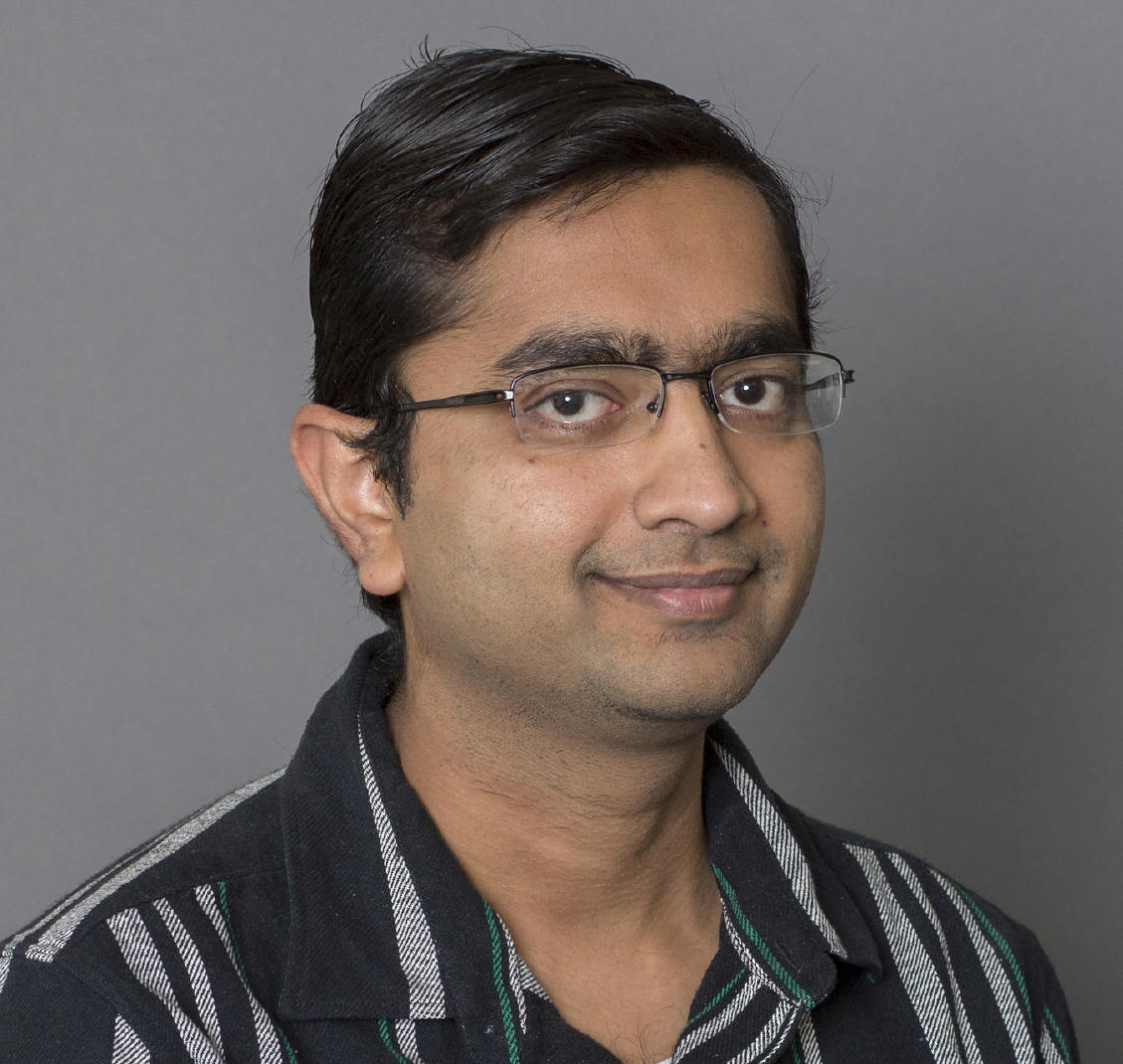}}]{Vaneet Aggarwal (S'08 - M'11 - SM’15)}
	received the B.Tech. degree in 2005 from the Indian Institute of Technology, Kanpur, India, and the M.A. and Ph.D. degrees in 2007 and 2010, respectively from Princeton University, Princeton, NJ, USA, all in Electrical Engineering.
	
	He is currently an Assistant Professor at Purdue University, West Lafayette, IN. Prior to this, he was a Senior Member of Technical Staff Research at AT\&T Labs-Research, NJ, and an Adjunct Assistant Professor at Columbia University, NY. His research interests are in applications of statistical, algebraic, and optimization techniques to distributed storage systems, machine learning,   and wireless systems. Dr. Aggarwal was the recipient of Princeton University's Porter Ogden Jacobus Honorific Fellowship in 2009. In addition, he received AT\&T Key Contributor award in 2013, AT\&T Vice President Excellence Award in 2012, and AT\&T Senior Vice President Excellence Award in 2014. 
\end{IEEEbiography}

\begin{IEEEbiography}[{\includegraphics[width=1in,height=1.25in,clip,keepaspectratio]{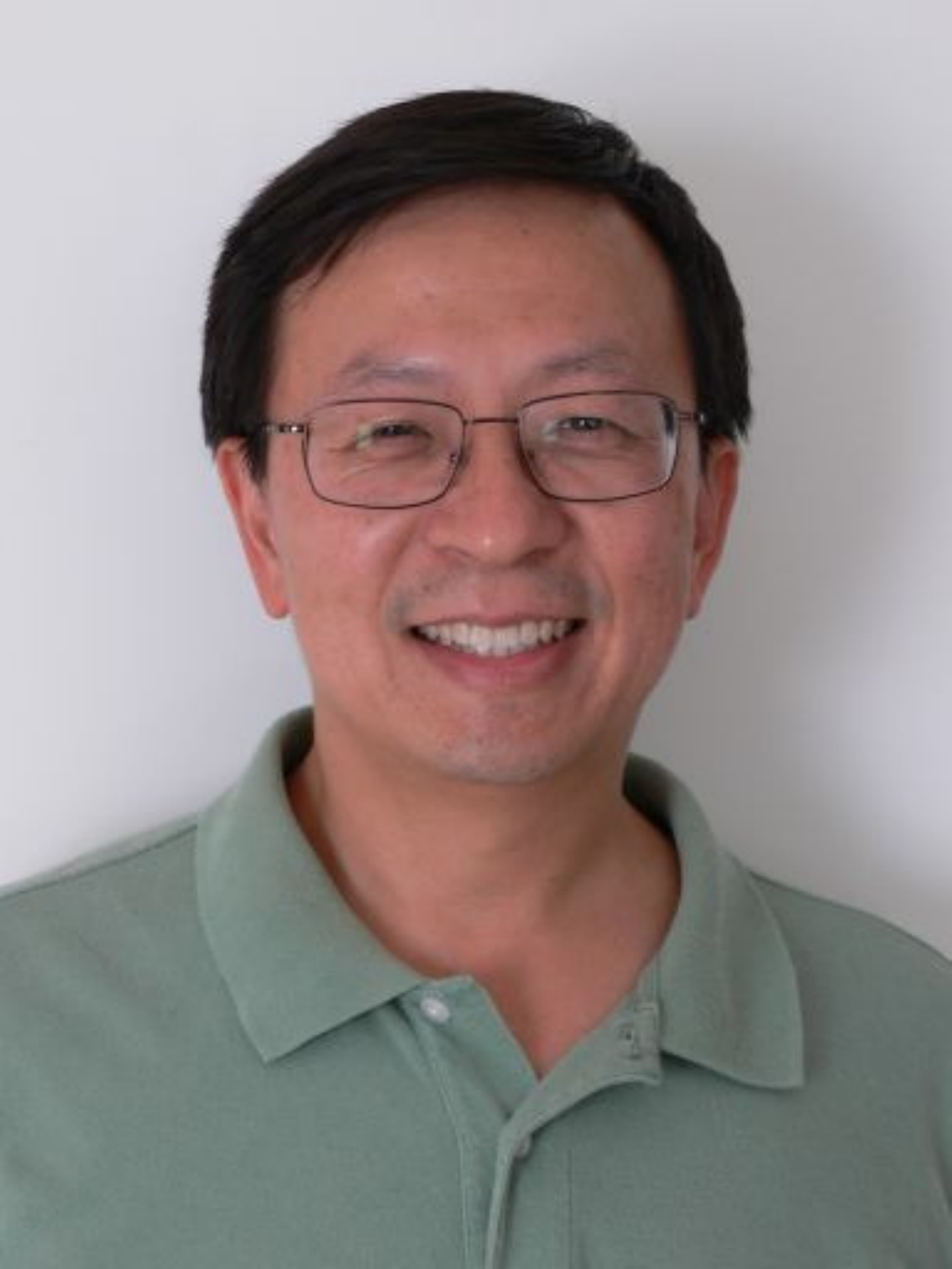}}]{Yih-Farn (Robin) Chen}
	is a Director of Inventive Science, leading the Cloud Platform Software Research department at AT\&T Labs  - Research.  His current research interests include cloud computing, software-defined storage, mobile computing, distributed systems, World Wide Web, and IPTV.  He holds a Ph.D. in Computer Science from University of California at Berkeley, an M.S. in Computer Science from University of Wisconsin, Madison,  and a B.S. in Electrical Engineering from National Taiwan University. Robin is an ACM Distinguished Scientist and a Vice Chair of the International World Wide Web Conferences Steering Committee (IW3C2).  He also serves on the editorial board of IEEE Internet Computing.
\end{IEEEbiography}

\begin{IEEEbiography}[{\includegraphics[width=1in,height=1.25in,clip,keepaspectratio]{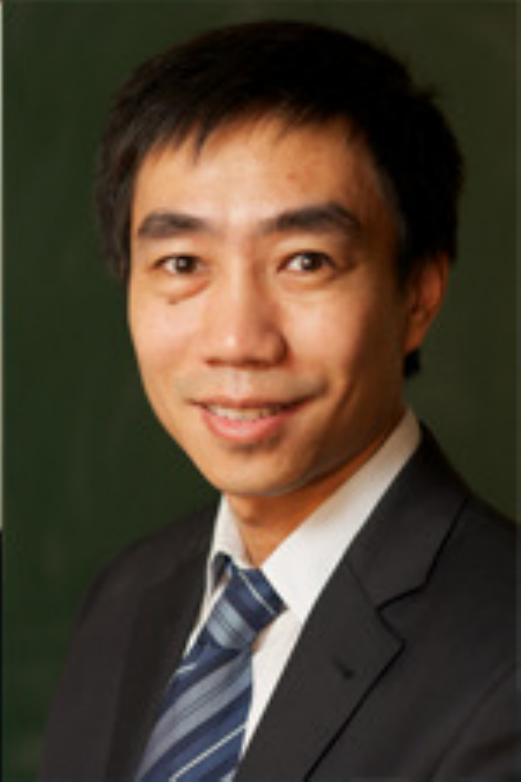}}]{Tian Lan (S'03-M'10)} received the B.A.Sc. degree from the Tsinghua University, China in 2003, the M.A.Sc. degree from the University of Toronto, Canada, in 2005, and the Ph.D. degree from the Princeton University in 2010. Dr. Lan is currently an Associate Professor of Electrical and Computer Engineering at the George Washington University. His research interests include cloud resource optimization, mobile networking, storage systems and cyber security. Dr. Lan received the 2008 IEEE Signal Processing Society Best Paper Award, the 2009 IEEE GLOBECOM Best Paper Award, and the 2012 INFOCOM Best Paper Award.
\end{IEEEbiography}

\begin{IEEEbiography}[{\includegraphics[width=1in,height=1.25in,clip,keepaspectratio]{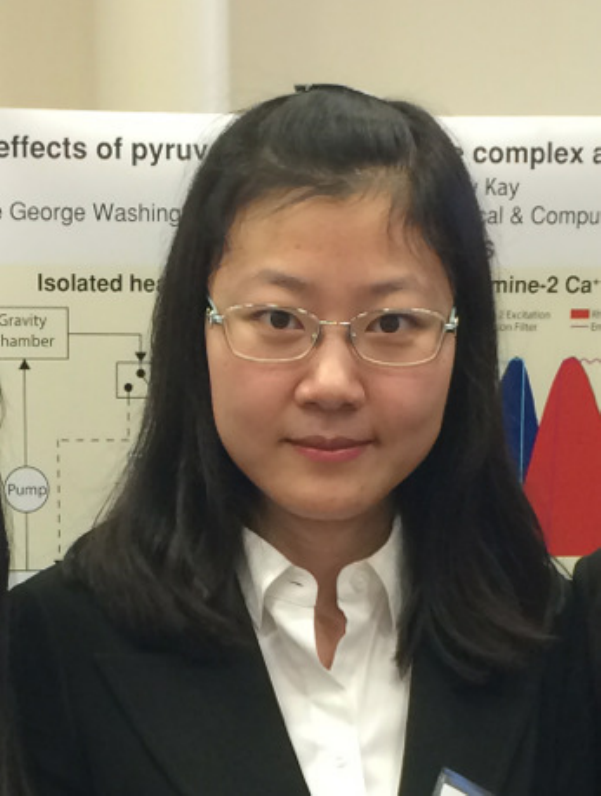}}]{Yu Xiang} received the B.A.Sc. degree from Harbin Institute of Technology in 2010, and the Ph.D. degree from George Washington University in 2015, both in Electrical Engineering. She is now a senior inventive scientist at AT\&T Labs-Research. Her current research interest are in cloud resource optimization, distributed storage systems and cloud storage charge-back.
\end{IEEEbiography}

%% file: testbed.tex
\section{Ceph Testbed}\label{testbed}

The underlying foundation of Ceph is the Reliable Autonomic Distributed Object Store (RADOS). Ceph provides three types of storage API's on top of RADOS: (i) Object Storage: the fundamental RADOS object-based storage system. (ii) Block Storage:  RADOS Block Device (RBD) that stores block device images as objects. (iii) Filesystem: this runs on top of the object storage with a metadata server that maps the directories and file names of the file system to objects stored within RADOS clusters. We use Ceph object storage in our testbed for our joint optimization algorithm with caching in erasure coded storage because i) Object storage is a good fit for large-scale cloud storage systems since it stores data more efficiently. ii) In terms of resilience, Ceph object storage provides both replicated and erasure-coded storage, while erasure-coded block storage on top of RADOS is not supported in Ceph yet. 

{\bf Ceph Object Storage Cluster:} A Ceph object storage cluster consists of two types of daemons: (i) Ceph monitor, which keeps a master copy of the cluster map and monitors the status of daemons in the cluster. (ii) Ceph object storage device(OSD), which is a physical or logical storage device used to store data as objects. Each OSD has two components: journal and data, during a write operation, it writes to the journal first and then writes data from journal to OSD data drive. Ceph stores data within pools, a pool consists of a bunch of OSDs, and Ceph users can specify which bunch of OSDs forms a certain pool through CRUSH rules. CRUSH is an algorithm Ceph used to pseudo-randomly store and retrieve data in OSDs with a uniform distribution of data across the cluster. CRUSH has a cluster map which contains a list of OSDs, and each OSD's physical location, and a list of rules that tell CRUSH how it should encode/replicate data in pools. To reduce the amount of metadata when storing data in OSDs, Ceph maps objects to placement groups, which are fragments of an object pool that place objects as a group into OSDs. The number of placement groups in an (n=k+m, k) erasure coded pool can be calculated as: 
\begin{equation}
\text{Number of Placement Groups}=\frac{\text{Number of OSDs}*100}{m}
\label{eq:npg}
\end{equation}

{\bf Erasure-coded Storage in Ceph:} When creating an erasure coded pool in Ceph, users need to specify erasure code parameters, the number of placement groups, OSDs belong to this pool, and failure domain of the pool (OSD, or host) etc., in an erasure-code profile. Ceph object storage is using {\em jerasure plugin} (a generic and flexible plugin that encapsulates the Jerasure code library) for default erasure code set up \cite{jerasure}. For an erasure-code profile with jerasure plugin, we specify erasure code with $k$, which is the number of data chunks, and $m$, which is the number of coded chunks, thus forming an  $(k+m,k)$ erasure code.

{\bf Ceph Cache Tier:} Ceph provides a cache tiering service to improve IO performance for storage devices. Users can create a Ceph cache tier as an overlay to the existing storage pool. Usually, the cache tier is a pool with faster but expensive storage devices (mostly SAS SSDs or Nvme SSDs). The existing back-end storage pool could be either an erasure-coded or replicated pool composed of slower and cheaper storage devices.  However, the cache tier pool has to be replicated as Ceph caching does not support erasure-coded storage yet. For a pool with cache tier, all IO traffic will be routed to the cache tier first. In the case of a cache miss, the data will be promoted from the storage tier to the cache tier. And the tiering agent will evict the least recently used (LRU) data from cache tier and flushes them to storage tier. 

%% file: apdx.tex
{\section{Proof of Lemma \ref{th:lemma_1}}\label{apdx}

The key idea of probabilistic scheduling is to get the file $i$ from set $\mathcal{A}_i$, which is chosen probabilistically, and thus is given by the parameters $\mathbb{P}(\mathcal{A}_i)$ $\forall \mathcal{A}_i\subseteq \mathcal{S}_i$ and $\forall i$, which involves $\sum_i {n_i\choose k_i}$ decision variables. While it appears prohibitive computationally, the authors of \cite{Yu_TON} demonstrated  that the optimization can be transformed into an equivalent form, which only requires $\sum_i n_i$ variables. The key idea is to show that it is sufficient to consider the conditional probability (denoted by $\pi_{i,j}$) of selecting a node $j$, given that a batch of $k_i$ chunk requests of file $i$ are dispatched. It is easy to see that for given $\mathbb{P}(\mathcal{A}_i)$, we can derive $\pi_{i,j}$ by
\vspace{-2mm}
\begin{eqnarray}
\pi_{i,j} = \sum_{\mathcal{A}_i:\mathcal{A}_i\subseteq \mathcal{S}_i} \mathbb{P}(\mathcal{A}_i) \cdot {\bf 1}_{\{ j\in\mathcal{A}_i\}}, \ \forall i \label{eq:pi}
\end{eqnarray}
where ${\bf 1}_{\{ j\in\mathcal{A}_i\}}$ is an indicator function which equals to 1 if node $j$ is selected by $\mathcal{A}_i$ and 0 otherwise. The equivalence proof of the probability over sets $\mathcal{A}_i$ and the nodes $j$ follows using  Farkas-Minkowski Theorem \cite{Angell:02}. 

Using this result, it is sufficient to study probabilistic scheduling via conditional probabilities $\pi_{i,j}$, which greatly simplifies the analysis. In particular, it is easy to verify that under our model, the arrival of chunk requests at node $j$ form a Poisson Process with rate $\Lambda_j=\sum_i \lambda_i\pi_{i,j}$, which is the superposition of $r$ Poisson processes each with rate $\lambda_i\pi_{i,j}$, $\mu_j$ is the service rate of node $j$. The resulting queuing system under probabilistic scheduling is stable if all local queues are stable.

Let ${\bf Q}_{max}$ be the maximum of waiting time $\{{\bf Q}_j, j\in \mathcal{A}_i\}  $. We first show that ${\bf Q}_{max}$ is upper bounded by the following inequality for arbitrary $z\in\mathbb{R}$:
\begin{eqnarray}
{\bf Q}_{max} \le z +\left[{\bf Q}_{\max} - z\right]^{+}  \le z + \sum_{j\in\mathcal{A}_i} \left[{\bf Q}_j - z\right]^{+}, \label{eq:lemma1_p1}
\end{eqnarray}
where $[a]^{+}=\max\{ a, 0\}$ is a truncate function. Now, taking the expectation on both sides of (\ref{eq:lemma1_p1}), we have
\begin{eqnarray}
& \mathbb{E}\left[{\bf Q}_{max} \right] & \le z +  \mathbb{E}\left[ \sum_{j\in\mathcal{A}_i} \left[{\bf Q}_j - z\right]^{+} \right] \nonumber \\
& & = z +  \mathbb{E}\left[ \sum_{j\in\mathcal{A}_i} \frac{1}{2} ( {\bf Q}_j - z + |{\bf Q}_j - z|) \right] \nonumber \\
& & =  z +    \mathbb{E}_{\mathcal{A}_i}\left[ \sum_{j\in\mathcal{A}_i} \frac{1}{2} ( \mathbb{E} [{\bf Q}_j] - z +  \mathbb{E}  |{\bf Q}_j - z|) \right], \nonumber \\
& &  =  z +  \sum_{j\in\mathcal{A}_i} \frac{\pi_{i,j}}{2} ( \mathbb{E} [{\bf Q}_j] - z +  \mathbb{E}  |{\bf Q}_j - z|),
\label{eq:lemma1_p2}
\end{eqnarray}
where $\mathbb{E}_{\mathcal{A}_i}$ denotes the expectation over randomly selected $k_i$ storage nodes in $\mathcal{A}_i \subseteq \mathcal{S}$ according to probabilities $\pi_{i,1},\ldots, \pi_{i,m}$. From Cauchy-Schwarz inequality, we have
\begin{eqnarray}
\mathbb{E}  |{\bf Q}_j - z| \le  \sqrt{ (\mathbb{E} [{\bf Z}_j] - z)^2 + {\rm Var}[{\bf Q}_j] } \label{eq:lemma1_p3} .
\end{eqnarray}
Combining (\ref{eq:lemma1_p2}) and (\ref{eq:lemma1_p3}), we obtain 
\begin{eqnarray}
\D & \bar{T}_i & \leq \min_{z\in \mathbb{R}} \left\{ z+\sum_{j\in \mathcal{S}_i} \frac{\pi_{i,j}}{2}  \left(\mathbb{E}[{\bf Q}_j] -z \right) \right.   \nonumber\\
& & \left.+ \sum_{j\in \mathcal{S}_i} \frac{\pi_{i,j}}{2} \left[ \sqrt{(\mathbb{E}[{\bf Q}_j]-z)^2+{\rm Var}[{\bf Q}_j]}\right] \right\}.
\end{eqnarray}

Using Pollaczek-Khinchin transform \cite{ISIT:12} to obtain the mean and variance of M/G/1 queueing delays, we obtain the result as in the statement of the Lemma.

}